%Paper: hep-th/9301088
%From: warner@faraday.usc.edu (Nicholas Philip Warner)
%Date: Thu, 21 Jan 93 16:58:44 GMT-0800

\input harvmac.tex
\def\LG{Lan\-dau-Ginz\-burg\ }
\def\coeff#1#2{{#1\over#2}}
\def\bar{\overline}

%
% referencing
\def\nup#1{{\it Nucl.\ Phys.} \ {\bf B#1\/}}
\def\ijmp#1{{\it Int.\ J. \ Mod. \ Phys.} \ {\bf A#1\/}}
\def\plt#1{{\it Phys.\ Lett.}\ {\bf#1\/}}
\def\cmp#1{{\it Commun.\ Math.\ Phys.} \ {\bf #1\/}}
\def\mpl#1{{\it Mod.\ Phys.\ Lett.} \   {\bf A#1}\ }
\def\prl#1{{\it Phys.\ Rev.\ Lett.\ } {\bf #1}\ }

\def\coeff#1#2{\relax{\textstyle {#1 \over #2}}\displaystyle}
% small coeff
\def\half{{1 \over 2}}
\def\inbar{\vrule height1.5ex width.4pt depth0pt}
\def\IC{\relax\,\hbox{$\inbar\kern-.3em{\rm C}$}}
\def\IP{\relax{\rm I\kern-.18em P}}
\def\IR{\relax{\rm I\kern-.18em R}}
\font\sanse=cmss12
\def\ZZ{\relax{\hbox{\sanse Z\kern-.42em Z}}}
\def\cR{{\cal R}}

%The following modify definitions that are made in Harvmac
\def\Titletwo#1#2#3#4#5{\nopagenumbers\abstractfont\hsize=\hstitle
\rightline{#1}\rightline{#2}
\vskip .8in\centerline{\titlefont #3}
\vskip .1in \centerline{\abstractfont {\titlefont #4} \footnote{*}{ #5}}
\abstractfont\vskip .5in\pageno=0}
\def\Date#1{\leftline{#1}\tenpoint\supereject\global\hsize=\hsbody%
\footline={\hss\tenrm\folio\hss}}% restores pagenumbers
%
% *****  End init.tex
% *********************************************************

\Titletwo{}{}
{ $N=2$ Supersymmetric Integrable Models}{and Topological Field Theories }{
Lectures given at the Summer School on High Energy Physics and Cosmology,
Trieste, Italy, June 15th -- July 3rd, 1992.  To appear in the proceedings.}
\centerline{N.P. Warner \footnote{**}{Work supported in part by the
DOE under grant No. DE-FG03-84ER40168, and also by a fellowship
from the Alfred P. Sloan foundation.}}
\bigskip \centerline{Physics Department}
\centerline{University of Southern California}
\centerline{Los Angeles, CA 90089-0484}
\vskip 2.0cm
These lectures review some of the basic properties of $N=2$ superconformal
field theories and the corresponding topological field theories.  One of my
basic aims is to show how the techniques of topological field theory can be
used to compute effective \LG potentials for perturbed $N=2$ superconformal
field theories.  In particular, I will briefly discuss the application of these
ideas to $N=2$ supersymmetric quantum integrable models.

\vskip 0.5cm
\centerline{\it Dedicated to the memmory of Brian Warr}
\vfill
\leftline{USC-93/001 }
\leftline{hep-th/9301088}
%
%\draft
\Date{January, 1993.}

\newsec{Introduction}
\nref\NPW{N.P. Warner, {\it ``Lectures on N=2 superconformal theories and
singularity theory'',in ``Superstrings '89,''} proceedings of the Trieste
Spring School, 3--14 April 1989.  Editors: M.\ Green, R.\ Iengo,  S.\
Randjbar-Daemi, E.\ Sezgin and A.\ Strominger.  World Scientific (1990).}
\nref\EMa{E.~Martinec, \plt{217B} (1989) 431.}
\nref\VW{C.~Vafa and N.P.~Warner, \plt{218B} (1989) 51.}
\nref\EMb{E.~Martinec, {\it ``Criticality, catastrophes and
compactifications,''}  V.G. Knizhnik memorial volume, L.~Brink {\it  et al.}
(editors): {\it Physics and mathematics of strings.} }
\nref\LVW{W.~Lerche, C.~Vafa and N.P.~Warner, \nup{324} (1989) 427.}
\nref\Gepner{D.~Gepner, \plt{222B} (1989) 207.}
\nref\HW{P.~Howe and P.~West, \plt{223B} (1989) 377.}
\nref\GVW{B.~Greene, C.~Vafa and N.P.~Warner,  \nup{324} (1989) 371.}
\nref\CGPA{S.~Cecotti,  L.~Girardello and A.~Pasquinucci, \nup{328} (1989) 701.
}

The study of two-dimensional $N=2$ supersymmetric field theories has shown
surprising longevity in an era when half-lives of research areas (and average
attention spans) are dropping well below a year.  There are probably several
reasons for the continued interest in $N=2$ supersymmetric theories, but I
believe that the most fundamental reason is that they have just the right
amount of supersymmetry.  They have enough supersymmetry so that they have
topological, and pseudo-topological, sectors whose quantum properties can be
computed semi-classically, and at the same time, these theories do not have so
much supersymmetry that their structure is so rigid as to render the theory
sterile and uninteresting.

There are now a number of very active areas of research in which $N=2$
supersymmetric field theories are finding interesting applications: these areas
include string theory, mirror symmetry, topological field theory, exactly
solvable lattice models, two dimensional theories of quantum gravity and
W-gravity, and even in polymer physics.  In these lectures my aim will be to
show how all the technology of perturbed $N=2$ superconformal field theories
and topological models provides a powerful set of tools in the analysis of
$N=2$ supersymmetric quantum integrable theories.  This approach to the subject
comes from a desire to mesh with the themes of this school, but also has the
virtue of getting to some physically interesting results without the
prerequisite of a course in string theory or algebraic geometry.  Moreover, I
will also be able to review a reasonable amount of the $N=2$ supersymmetry
technology that is currently finding applications elsewhere.  Consequently,
whenever possible I will try to indicate where my lectures connect, albeit
tangentially, with the other currently active fields of research involving
$N=2$ supersymmmetry.  I will also attempt to make my lectures relatively
self-contained by reviewing the basic ideas of $N=2$ superconformal theories,
but this review will be somewhat brief and more details may be found in my
lectures at an earlier school at the ICTP \NPW\ or in the earlier papers
\refs{\EMa {--} \CGPA}.

The topics that I will cover here are:

\item{(i)}  $N=2$ superconformal field theories, chiral rings and effective \LG
potentials.
\item{(ii)}  Topologically twisted $N=2$ superconformal field theories
\item{(iii)}  Perturbed $N=2$ superconformal field theories, both topological
and non-topological.
\item{(iv)} Effective \LG potentials and kink masses in perturbed $N=2$
superconformal field theories.
\item{(v)}  Computing effective \LG potentials using topological field theory.
\item{(vi)}  Simple $N=2$ supersymmetric quantum integrable models and their
soliton structure.

The lectures of Dennis Nemeschansky will start where I finish:  he will discuss
soliton scattering matrices in the $N=2$ supersymmetric quantum integrable
models.  Cumrun Vafa will start his lectures by showing how a number of the
concepts that I introduce for $N=2$ superconformal field theories can be easily
generalized to massive $N=2$ supersymmetric theories, and he will then show how
 the topological structure of these massive models can be used to determine
much about the ``pseudo-topological'' sectors of the theory.

\newsec{$N=2$ Superconformal Field Theories.}

\subsec{The operators and primary fields}

In an $N=2$ superconformal field theory the energy momentum tensor, $T(z)$, is
supplemented by two supercharges, $G^+(z)$ and $G^-(z)$, and a $U(1)$ current,
$J(z)$.  These four generators have conformal weights $2$, $\coeff32$,
$\coeff32$ and $1$ respectively, and have operator product expansions:
\eqn\SCAOPE{\eqalign{
G^\pm(z) G^\mp(w) &~=~ {{{2 \over 3} c } \over {(z - w)^3}} ~\pm~ {{2J(w)}
\over {(z - w)^2}} ~+~ {{2T(w) \pm \partial_w J(w)} \over {(z - w)}}+ \dots \cr
J(z) G^\pm (w) &~=~ \pm~ {{G^\pm (w)} \over {(z - w)}} + \dots \cr
J(z) J(w) &~=~ {{{1 \over 3}c} \over {(z - w)^2}}+ \dots \cr
T(z) J(w) &~=~ {{J(w)} \over {(z - w)^2}} ~+~ {{\partial_w J(w)} \over
{(z - w)}}+ \dots \cr
T(z) G^\pm (w) &~=~ {{{3 \over 2} G^\pm (w)} \over {(z - w)^2}} ~+~
{{\partial_w G^\pm (w)} \over {(z - w)}}+ \dots \cr
T(z) T(w) &~=~ {{{1 \over 2} c} \over {(z - w)^4}} ~+~ {{2 T(w)} \over {(z
- w)^2}} ~+~ {{\partial_w T(w)} \over {(z - w)}}+ \dots~, \cr}}
where, as usual, $+ \dots~$ means plus terms that are finite in the limit
as $z \rightarrow w$.  Note the presence  of the combination ${2T(w) \pm
\partial_w J(w)}$ in the operator product $G^\pm(z) G^\mp(w)$.  This will be
important in the subsequent discussion of topologically twisted theories.  One
can pass to modes and write:
\eqn\SCAops{\eqalign{
T(z) &~=~ \sum_{n = - \infty}^\infty L_n~ z^{-n-2} \cr
G^\pm (z) &~=~ \sum_{n = - \infty}^\infty G_{n \pm a}^\pm~ z^{-(n \pm a) -
3/2} \cr
J(z) &~=~ \sum_{n = -\infty}^\infty J_n ~z^{-n-1}. \cr}}
The parameter $a$ is a real number, and determines the branch cut properties of
$G^\pm(z)$.  A field theory with $N=2$ superconformal symmetry with $a=0$ is
usually said to be in a Ramond sector, and if the theory has $a=\half$ then it
is said to be in a Neveu-Schwarz (NS) sector.  I will simplify my life here by
working almost entirely with theories in the NS sector.  As I will describe
later, it is very simple to convert results obtained in the NS sector into
results for the superalgebra for any value of $a$.

In terms of modes, the foregoing operator products can be written as:
\eqn\SCA{\eqalign{
[L_m, L_n] &~=~ (m-n) L_{m+n} + {c \over {12}} m(m^2-1)\delta_{m+n,0}\cr
[J_m,J_n] &~=~ {c \over 3} m \delta_{m+n,0}\cr
[L_n,J_m] &~=~ -m J_{m+n}\cr
[L_n, G_{m\pm a}^\pm] &~=~ \left( {n \over 2} - (m \pm a)\right) G_{m+n \pm
a}^\pm \cr
[J_n, G_{m\pm a}^\pm] &~=~ \pm~ G_{m+n \pm a}^\pm \cr
\{G_{n+a}^+ , G_{m-a}^- \} &~=~ 2L_{m+n} + (n-m+2a) J_{n+m} + {c \over
3} \left[ (n+a)^2 - {1 \over 4} \right] \delta_{m+n,0}~,\cr}}
where $m$ and $n$ are integers.

One should remember that in a conformal field theory there is both a
holomorphic (left-moving) and an anti-holomorphic (right-moving) sector, and
these two sectors have to be combined in the complete theory.  Much of the time
I will suppress the discussion of the anti-holomorphic sector, but throughout
these lectures I will implicity require that its structure be directly parallel
to the structure of the holomorphic sector.  In particular, this means that I
will assume that the anti-holomorphic sector has $N=2$ superconformal symmetry
with generators \foot{Objects with a tilde, $\widetilde{\ } $, will generically
denote anti-holomorphic counterparts of holomorphic quantities.}: $\widetilde
T(\bar z)$, $\widetilde G^+(\bar z)$, $\widetilde G^-(\bar z)$ and  $\widetilde
J(\bar z)$.

A primary field, $\psi(z)$, of the $N=2$ superconformal algebra satisfies:
\eqn\pfield{\eqalign{
T(z)~ \psi(w) &~=~ {h \over {(z - w)^2}}~ \psi (w) ~+~ {{\partial_w \Psi
(w)}\over {( z - w)}} ~+~\dots \cr
J(z)~ \psi(w) &~=~ {q \over {(z-w)}}~ \psi (w) ~+~\dots \cr
{G^\pm} (z)~ \psi (w) &~=~ {1 \over {(z-w)}}~ {\Lambda^\pm}(w)
{}~+~\dots \ ,\cr}}
where the fields $\Lambda^\pm (w)$ are the super-partners of $\psi(w)$.  In
terms of states and modes, the foregoing is equivalent to:
\eqn\prim{\eqalign{G_r^\pm~ | \psi \! > ~&=~ 0\ , \quad r \geq \half \ ;
\qquad L_n~ | \psi\! > ~=~ J_n~ | \psi \!> ~=~ 0 , \quad n \geq 1\ ; \cr
G_{-\half}^\pm~ | \psi \! >  ~&=~  | \Lambda^\pm \! >\ , \qquad L_0 ~
| \psi \!> \, ~=~ h | \psi \!>\ , \qquad J_0~ | \psi \! >
\, ~=~ q | \psi \! > \  .}}
In order to obtain unitary representations of this algebra one must have
\ref\BFK{W.~Boucher, D.~Friedan and A.~Kent, \plt{172B} (1986) 316.}:
$c \geq 3$ or $c = 3 -{6 \over {(k+2)}}$; $k=1,2, \dots$.  The models with
$c =3k/(k+2)$ are called the $N=2$ superconformal minimal models.  For these
minimal models there are only finitely many highest weight irreducible
representations.  These irreducible representations are determined by the
conformal weight, $h_0$, and $U(1)$ charge, $q_0$, of the ground state $| h_0,
q_0 \! >$, and the allowed values of $h_0$ and  $q_0$ (in the NS sector) are:
\eqn\minmodqns{h_0 ~=~ {{\ell(\ell+2) - m^2} \over {4(k+2)}} \ , \qquad
q_0 ~=~ {m \over (k+2)} \ ,}
for $\ell = 0,1,2, \dots ,k$ and $m = -\ell, -\ell + 2, \dots, \ell -2, \ell$.
The remaining states in the representation are then obtained from the ground
state by acting with $L_{-n}$, $J_{-n}$ and $G_{-r}^+$ for $n, r > 0$.  Because
of this simple structure, these minimal models will be used to give examples
throughout my lectures.  There are, of course, considerably more $N=2$
superconformal theories that are fairly well understood, for example, there are
the $N=2$ superconformal coset models \ref\KS{ Y.\ Kazama and H.\ Suzuki,
\plt{216B} (1989) 112; \nup{321} (1989) 232.} and their closely related kin,
the $N=2$ super-$W$ algebras  \nref\KIto{K.~Ito, \plt{259B} (1991) 73.}
\nref\HN{D.~Nemeschansky and S.~Yankielowicz,{ \it ``$N=2$ W-algebras,
Kazama-Suzuki Models and Drinfeld-Sokolov Reduction,''} USC-preprint USC-007-91
(1991)} \nref\Romans{L.J.~Romans, \nup{369} (1992) 403.} $\!$\refs{\KIto {--}
\Romans}.

\subsec{The chiral ring}

Consider a general $N=2$ superconformal theory.  A field $\phi(z)$ will be
called {\it chiral} if it satisfies:
\eqn\chiralfld{\left( G^+_{-\half} \phi \right)(z) ~\equiv~ 0\ .}
That is, half of its super-partners vanish.  A field is called chiral, primary
if it is both chiral and primary.  The set of such fields is called the {\it
chiral ring} and will be denoted by $\cR$.  As we will see, it does indeed have
a ring structure.

To derive the properties of chiral primary fields one first considers unitarity
bounds.  Suppose that  $| \psi \!>$ is any state.  Then because
$(G_r^\pm)^\dagger = G_{-r}^\mp$, one has:
\eqn\unitbnda{0 \ ~\leq~\ \  <\! \psi | \{ G_{-\half}^\pm \, , \ G_{\half}^\mp
\} | \psi \! >  \ \ ~=~ \ \  < \! \psi |~ 2 h_\psi \mp q_\psi~ | \psi \! > \ ,}
and hence for all states in a (unitary) $N=2$ superconformal theory one has
\eqn\unitbndb{h ~\ge ~ \half \vert q \vert\ .}
{}From the bound \unitbnda\ one can show that an equivalent characterization of
$\cR$ is that it is precisely the set of fields that have $ q >0$ and saturate
the bound in \unitbndb, {\it i.e.} for which one has \foot{ The situation is
not so simple for non-unitary theories, see for example
\ref\BLNW{M.~Bershadsky, W.~Lerche,  D.~Nemeschansky and N.P.~Warner {\it
``Extended $N=2$ Superconformal Structure of Gravity and $W$-Gravity Coupled to
Matter,''}  Caltech preprint CALT-68-1832, CERN preprint CERN-TH.6694/92,
Harvard preprint HUTP-A061/92, USC preprint USC-92/021.}.  Here I will only
consider unitary theories.}
\eqn\cringdefb{ h ~=~ \half q \ .}
This may be seen by first observing that if $h_\psi = \half q_\psi$ then
\unitbnda\ immediately implies $G_{-\half}^+ | \psi \! > = $ $G_{\half}^- |
\psi \! > = 0$.  In addition, all the states $G_{r}^+ | \psi \! >$, $G_{r +
1}^- | \psi \! >$,  $L_n | \psi \! >$ and $J_n | \psi \! >$ for $n,r > 0$ must
be zero since they would otherwise violate \unitbndb.  Therefore such a state
$|\psi \! >$ must be both chiral and primary.

It is now easy to see how $\cR$ inherits a ring structure.  Consider the
operator product of two chiral primary fields $\phi_i$ ($i = 1,2$) of conformal
weight $h_i$ and charge $q_i = 2h_i$.  Suppose that $\psi$ is some operator
that appears on the right hand side  of the operator product:
$$
\phi_1(z) \ \phi_2(w) ~=~  \dots +~ (z-w)^{h_\psi - h_1 - h_2} \psi(w) ~+\dots
$$
Then charge conservation, \unitbndb\ and \cringdefb\ imply that the power,
$h_\psi - h_1 - h_2$, is non-negative and vanishes if and only if $h_\psi =
\half q_\psi$, that is, it vanishes if and only if $\psi$ is both chiral and
primary.  Therefore the following is a well defined (associative and
commutative) multiplication that closes into $\cR$:
\eqn\multpdefn{ (\phi_1 \cdot  \phi_2)(w) ~=~ \lim_{z \to w}{\ \phi_1(z) \
\phi_2(w)} \ .}
The associativity and commutativity of this product follow trivially from the
properties of the operator product.

Because it is associative and commutative, the ring, $\cR$, may be thought of
as a polynomial ring.  There will be generating fields $x_a \in \cR$ such that
\eqn\ReqlpJ{ \cR ~ = ~ {\cal P}[x_a] /{\cal J} \ , }
where ${\cal P}[x_a]$ is the ring of complex polynomials in the $x_a$ and
${\cal J}$ are the ``vanishing relations''.  That is, ${\cal J}$ is an ideal of
${\cal P}[x_a]$  consisting of all the polynomials in $x_a$ that vanish in the
product defined above.

The simplest examples of chiral rings are obtained from the minimal models.
The primary fields are labelled $\Phi^\ell_m$ and have conformal weight and
$U(1)$ charge given by \minmodqns.  One then has $h = \half q$ if and only if
$m=\ell$.  Let $ x = \Phi_1^1$, and then one has $\Phi_\ell^\ell = x^\ell$ in
the obvious sense.  However, recall that one has $\ell = 0,1, \dots , k$, and
so one must have $x^{k+p} \equiv 0$ for $p \ge 1$.  Consequently, the ideal,
${\cal J}$, is precisely generated by $x^{k+1}$, {\it i.e.} ${\cal J}= {\cal
P}[x]\{x^{k+1}\}$, and $\cR$ has a basis $\{1,x,x^2,$ $ \dots, x^k\}$.

\subsec{Further properties of the chiral ring}

Conjugate to the chiral ring, $\cR$, is the anti-chiral ring, $\bar \cR$.  This
simply consists of all operators that are primary and anti-chiral, that is,
primary and annihilated by $G_{-\half}^-$.  The structure of $\bar \cR$ is
directly parallel to that of the chiral ring.  It can also be characterized as
consisting of all the fields that satisfy $h = -\half q$.

There is another basic tool in the analysis of $N=2$ superconformal field
theories and that is {\it spectral flow}.  The basic idea is that in any theory
containing a $U(1)$ current, it is an elementary operation to shift the $U(1)$
charge of any operator.  Specifically, in the $N=2$ superconformal theory one
can write:
\eqn\uonebose{J(z) ~=~ i \sqrt{{c \over 3}} ~\partial X(z) \ ,}
where $X(z)$ is a canonically normalized boson.  One then introduces an
operator ${\cal U}_\theta (z)$ defined by:
\eqn\Udefn{{\cal U}_\theta (z) ~=~ exp\left[ i \sqrt{{c \over 3}}~\theta ~X(z)
\right] \ .}
Let $\psi(z)$ be some operator of charge $q$, then after spectral flow by
$\theta$ one obtains an operator, $\psi_\theta(z)$, of charge $q + \theta {c
\over 3}$, defined by
\eqn\flowpsi{\psi_\theta (w) ~\equiv ~ \lim_{z \to w}~(z-w)^{-q \theta } ~
{\cal U}_\theta (z) ~ \psi(w) \ .}
I will denote the corresponding mapping on operators by ${\cal U}_\theta$.
There are several uses for this map.  First of all, ${\cal U}_\theta$ maps the
NS representation of the $N=2$ superconformal algebra onto a representation
with the parameter $a$ of \SCAops\ and \SCA\ given by $a = \theta $.  Putting
it slightly differently, if one uses spectral flow to conjugate the operators
in the $N=2$ superconformal algebra, then they change according to:
\eqn\specflow{\eqalign{{\cal U}_\theta  L_n {\cal U}_\theta^{-1} &~=~
L_n - \theta J_n+{c\over 6}\theta^2 \delta_{n,0} \cr
{\cal U}_\theta J_n {\cal U}_\theta^{-1} &~=~
J_n - {c\over 3}\theta \delta_{n,0}  \cr
{\cal U}_\theta G^+_r {\cal U}_\theta^{-1} &~=~ G^+_{r-\theta} \cr
{\cal U}_\theta G^-_r{\cal U}_\theta^{-1}&~=~G^-_{r+\theta}.} }
If $\theta$ is an integer then the algebra maps back into itself.  One can also
verify that if $\theta = -1$ then $\cR$ maps one-to-one and onto $\bar \cR$,
and if $\theta = +1$ then $\bar \cR$ maps one-to-one and onto $\cR$.

Consider the operators $\rho(z)$ and $\bar \rho(z)$ defined by:
\eqn\rhodefn{\eqalign{\rho(z) ~ &\equiv~  e^{i \sqrt{{c \over 3}} ~X(z)} ~=~
{\cal U}_1 (z) \cr \bar \rho(z) ~ &\equiv~  e^{-i \sqrt{{c \over 3}} ~X(z)} ~=~
{\cal U}_{-1} (z)\ .}}
These may be viewed as spectral flows of the vacuum state by one unit.  It is
trivial to check that $\rho$ and $\bar \rho$ are elements of $\cR$ and $\bar
\cR$ respectively, and they satisfy $h = \pm \half q = c/6$.  It is also easy
to see that these states are the unique states in the theory satisfying this
equation, since any such state under spectral flow can be taken to a state with
$h=0$ and $q=0$, which must be the vacuum.  The states $\rho$ and $\bar \rho$
are also the (unique) elements of maximal dimension in $\cR$ and $\bar \cR$.
This follow from the inequality
\eqn\hbound{0 ~ \ \leq~\ \ <\! \phi | \{ G_{-{3 \over 2}}^\pm \, , \ G_{{3
\over 2}}^\mp \} | \phi \! >\ \  ~=~ \ \ < \! \phi |~ 2 h_\phi \mp 3 q_\phi~ +
\coeff23 c | \phi \! >\ ,}
which implies that for $h \le c/6$ for all elements of $\cR$ and $\bar \cR$.
It is also useful to note that this inequality also implies:
\eqn\Grho{G_{-{3 \over 2}}^+ ~\rho ~=~ G_{-{3 \over 2}}^- ~\bar \rho
\equiv 0 \ ,}
which also follows from the fact that $\rho$ and $\bar \rho$ are spectral flows
of the vacuum.

Another important use of spectral flow is to note that for $\theta = \pm \half$
the spectral flow maps the NS sector into the Ramond sector and vice-versa
\foot{In a string theory this operation corresponds to space-time
supersymmetry.}.  From \specflow\ one sees that under such a spectral flow
$(L_0 \pm \half J_0)_{\rm NS}$ $\to (L_0 - {c \over 24})_{\rm Ramond}$.
Consequently, under the appropriate spectral flow, $\cR$ or $\bar \cR$ map to
the Ramond ground states.  This characterization of chiral primary fields makes
them easier to compute in practice \LVW, but the Ramond ground states do not
exhibit the ring structure in the simple manner that is evident for their NS
counterparts.  From the Ramond characterization we also see that the chiral
ring is finite dimensional, since, in any unitary theory the degeneracy at any
energy level is finite.

The similarity between chiral rings and cohomology of differential forms may
already be apparent.  This parallel can be made even more explicit by
establishing a ``Hodge decomposition theorem,'' which says that any state,
$\psi$, can be written in the form:
\eqn\hodge{ | \psi \! > ~=~ | \phi \! > ~+~ G_{-\half}^+ | \chi_1 \! >  ~+~
G_{\half}^- | \chi_2 \! > \ ,}
where $| \phi \! >$ is a chiral primary, and $| \chi_1 \! >$ and $| \chi_2 \!
>$ are some other states.  Moreover, if $\psi$ is itself chiral, {\it i.e.}
$G_{-\half}^+ \psi = 0$, then one can take $| \chi_2 \! > = 0$.  This is
elementary to to prove using a Rayleigh-Ritz method:  one considers states, $|
\chi \! >$,  of the  form $| \chi \! > ~=~ | \psi \! >$ $~-~ G_{-\half}^+ |
\chi_1 \! > $  $ ~-~ G_{\half}^- | \chi_2 \! > $  and chooses $| \chi_1 \! >$
and $| \chi_2 \! >$ so as to minimize the norm of $\chi$.  It then follows that
$| \chi \! >$ is chiral and primary.  If $\psi$ is chiral, then take the inner
product of both sides of \hodge\ with $<\! \chi_2 | G_{-\half}^+ $, and one
then sees that $G_{\half}^- | \chi_2 \! > $ must be zero.

Since one has $\big(G_{-\half}^+\big)^2 = 0$, and chiral fields may all be
written in the form $| \phi \! >+G_{-\half}^+ | \chi_1 \! >$, where $| \phi \!
>$ is a chiral primary, it follows that $\cR$ is precisely the cohomology of
$G_{-\half}^+$.  In several situations the operator $G_{-\half}^+$ reduces to a
more familiar cohomological operator.  On coset conformal field theories it
becomes a loop-space Lie algebra cohomology operator, and on Calabi-Yau
manifolds $G_{-\half}^+$ becomes one of the Dolbeault operators \foot{Indeed ,
one can think of $G_{-\half}^+$ and $G_{\half}^-$, and their anti-holomorphic
counterparts $\widetilde G_{-\half}^+$ and $\widetilde G_{\half}^-$, as being
conformal field theoretic generalizations of the Dolbeault operators
$\partial$, $\delta$, $\bar \partial$ and $\bar \delta$.}.

So far I have only  discussed the holomorphic sector of the $N=2$
superconformal field theory.    There are also chiral and anti-chiral rings in
the anti-holomorphic sector.  There are thus four choices of ring: $(c,c)$,
$(c,a)$, $(a,c)$ and $(a,a)$, where $c$ and $a$ denote chiral and anti-chiral
respectively, and the entries in $(\ , \ )$ denote the holomorphic and
anti-holomorphic sectors.   The $(c,c)$ and $(a,a)$ rings are complex
conjugates of each other, as are the $(c,a)$ and $(a,c)$ rings.  However, in a
given $N=2$ superconformal theory, the two rings $(c,c)$ and $(c,a)$ are
distinct and frequently completely different.  From the computational point of
view in an $N=2$ superconformal theory it is easy to pass from one ring to the
other: one simply reverses the sign of the anti-holomorphic $N=2$, $U(1)$
current.  However, if the $N=2$ superconformal theory has a geometric origin,
such as coming from a compactification on a Calabi-Yau manifold, then the two
rings can have extremely different origins.  The conformal field theory then
puts on the same footing, two rings that are radically different from the point
of view of algebraic geometry.  This observation is the origin of mirror
symmetry in Calabi-Yau manifolds \nref\Dixon{L.~Dixon, {\it Some World-Sheet
Properties of Superstring Compactifications, on Orbifolds and Otherwise},
Lectures given at the 1987 ICTP Summer Workshop in High Energy Phsyics
and Cosmology, Trieste, Italy, Jun 29 -- Aug 7, 1987, in {\it Superstrings,
Unified Theories and Cosmology 1987}, G.~Furlan {\it et al.} editors, World
Scientific, (1988).}\refs{\Dixon,\LVW}:  If one can find a Calabi-Yau manifold,
${\cal M}$,  that gives rise to a particular $N=2$ superconformal field theory,
with the $(c,c)$ and $(c,a)$ rings each having a particular geometric origin,
then one should be able to find a manifold, $\widetilde {\cal M}$, that gives
rise to exactly the same $N=2$ superconformal theory but with the geometric
origins of the $(c,c)$ and $(c,a)$ rings inverted.  The fact that this is
possible has revolutionized an area of algebraic geometry. A recent review of
the subject may be found in \ref\mir{{\it ``Essays on Mirror Symmetry,''}
edited by S.-T. Yau. }.

Returning to superconformal theories, from now on when I refer to the chiral
ring of a complete $N=2$ superconformal theory I will generally mean the
$(c,c)$ ring, and I will restrict myself, for simplicity, to scalar chiral
primary fields.

\subsec{Landau-Ginzburg formulations of $N=2$ superconformal theories.}

The poor man's definition of when a $N=2$ superconformal theory has a \LG
formulation is that there must be a single quasihomogeneous function, $W_0$,
that characterizes the chiral ring in the following manner.  The ring, $\cR$,
has generators, $x_a$ of conformal weights  $h_a = \widetilde h_a$  $ = \half
\omega_a$, and $W_0$ is a function of these $x_a$'s having the quasihomogeneous
scaling property:
\eqn\quasih{W_0(\lambda^{\omega_a}~ x_a) ~=~ \lambda~W_0(x_a)\ ;}
and the ring itself must be given by:
\eqn\ReqLG{ \cR ~ = ~ {{\cal P}[x_a] \over \Big\{{\partial W_0 \over \partial
x_a} \Big \} } \ , }
where ${\cal P}[x_a]$ is the ring of polnomials in $x_a$ and $\big\{{\partial
W_0 \over \partial x_a}\big \}$ denotes the ideal generated by the partial
derivatives of $W_0$.  As an example, the chiral ring of the minimal models has
a \LG potential: $W_0(x) = x^{k+2}$, where $x \equiv \Phi^1_1$.

\nref\CGPB{S.\ Cecotti,  L.\ Girardello and A. Pasquinucci, \ijmp{6} (1991)
2427. }
The foregoing ``definition'' obscures the basic, and important physics that is
really required of a \LG formulation.   So to repair this omission I will
summarize the basic idea, further details can be found in
\refs{\NPW{--}\EMb,\HW,\CGPA,\CGPB}.  To say that an $N=2$ superconformal
theory has a \LG formulation really means that one can obtain it from an $N=2$
supersymmetric field theory that has a superpotential $W(x_a)$.  One then looks
for infra-red fixed points of the renormalization group flow.  According to
well substantiated folklore there are non-renormalization theorems that mean
that $W(x_a)$ only scales through wave-function renormalization, and at the
fixed point  this superpotential must scale to a superpotential, $W_0$, with
the quasihomogeneous scaling property \quasih\ where $\omega_a$ is the scaling
dimension ($h_a + \widetilde h_a$) of the scalar field $x_a$.  In this field
theory, the fields $x_a$ (and polynomials in them)  are defined precisely so as
 to be chiral in the supersymmetric sense.  At a fixed point of the
renormalization group flow one can also use the superpotential to determine
which of these polynomials in the $x_a$ are primary.  Having an effective \LG
superpotential, $W_0$, means precisely that the partial derivatives ({\it i.e.}
variations) of it must, via field equations, be proportional to
superderivatives ($G_{-\half}^+$) of something.  Conversely, if any polynomial
in $x_a$ is given by $G_{-\half}^+$ acting on something else then this fact
must be derivable from an effective field equation.  Thus the right-hand side
of \ReqLG\ characterizes all the chiral fields in the theory modulo chiral
fields that are given by $G_{-\half}^+$ acting on something else; this is
precisely the chiral ring.  It follows from the general properties that we have
already established about $N=2$ superconformal theories that the chiral ring is
a finite polynomial ring in which all the fields have their naive scaling
dimensions.  It is also worth mentioning that it is an elementary result of
singularity theory \nref\Arnbk{V.I. Arnold, {\it Singularity Theory}, London
Mathematical Lecture Notes Series: 53, Cambridge University Press (1981); V.I.
Arnold, S.M. Gusein-Zade and A.N. Varchenko, {\it Singularities of
Differentiable Maps}, volumes 1 and 2, Birkh\"auser (1985).} \nref\Arn{V.I.
Arnold, {\it Russian Mathematical Surveys,} {\bf 28} (1973) 19; {\bf 29} (1974)
11; {\bf 30} (1975) 1.}\nref\MBG{M.B. Green, another gratuitous
reference.}$\!\!\!$ \refs{\Arnbk{--}\MBG} that the polynomial:
\eqn\hessian{H(x_a) ~=~ det \left( {\partial^2 W_0(x_a) \over \partial x_b
\partial x_c } \right) }
is the unique element of maximal dimension in the ring defined by \ReqLG.  The
scaling dimension of $H(x_a)$ is easily seen to be $\sum_a (1 - 2 \omega_a)$.
Since $H(x_a)$ is maximal, it must be identified with $\rho$, whose dimension
is $c/3$, and as a result we see that we must have:
\eqn\cvalue{c ~=~ 3 ~ \sum_a (1 - 2 \omega_a)\ .}
\nref\OW{E.~Witten and D.~Olive, \plt{78} (1978) 97.}
\nref\FMVW{P.~Fendley, S.~Mathur, C.~Vafa and N.P.~Warner, \plt{243B} (1990)
257.}
It will be of importance later to note that if a supersymmetric theory has a
superpotential $W(x_a)$, then the effective bosonic potential is given by
$|\nabla W |^2$.  This means that the extremal points of $W$ correspond to zero
energy ground states of the supersymmetric field theory.  At the infra-red
fixed point of the renormalization group flow all of these ground states come
together making a multi-critical point.  It is easy to establish that the
number of such critical points is the Witten index of the theory and is equal
to the number of Ramond ground states in the conformal theory
\refs{\OW,\VW,\FMVW}.

A good analogy is to consider the $c=\half$ Virasoro minimal model.  This
appears at the infra-red fixed point of the \LG description of the Ising model.
 In these lectures I am considering $N=2$ supersymmetric generalizations, and
because of the remarkable properties of $N=2$ superconformal theories, and the
non-renormalization theorems of the $N=2$ supersymmetric theories, we can get
an exact quantum effective potential that gives us exact information about the
theory at (and near) the conformal point.  This fact is completely contrary to
one's experience with the two-dimensional Ising model, for which the \LG
description becomes only barely qualitative near the conformal point.

This parallel with the Ising model, and the labelling of this formulation of
certain $N=2$ superconformal models with title Lan\-dau-Ginz\-burg, raises the
fundamental question of whether such $N=2$ superconformal models can be
obtained from statistical mechanical systems or lattice models.  Such
connections with statistical mechanics were unclear when the the \LG
formulation of $N=2$ superconformal models was first introduced, however it has
recently been shown \ref\MNW{Z.~Maassarani, D.~Nemeschansky and N.P.~Warner,
{\it ``Lattice Analogues of $N=2$ Superconformal Models via Quantum Group
Truncation,''}  USC preprint USC-92/007, to appear in {\it Nucl.~Phys.} ~B.}
how to formulate a broad class of exactly solvable lattice models whose
continuum limit at the critical temperature are precisely the $N=2$
superconformal (hermitian, symmetric) coset models of \KS.  These models
include all the $N=2$ superconformal coset models that are known to have a \LG
formulation \foot{There are highly non-trivial infinite series of, and several
sporadic, coset models that have \LG formulations.  There are also infinitely
many coset models that do not have \LG formulations.}.  Moreover, the natural
order parameters of these lattice models renormalize to the chiral primary
fields at the conformal point.  Thus an underlying original hope has been
realized: the scalar chiral primaries are indeed \LG fields in the sense of
being order parameters of some statistical mechanical system.

\newsec{Twisted $N=2$ supersymmetric theories}

\subsec{The topological matter models}

It follows from Witten's original work on topological field theories
\ref\wittop{E.~Witten, \cmp{117} (1988) 353; \cmp{118} (1988) 411;
\nup340 (1990) 281.} that one can twist $N=2$ superconformal models and obtain
a topological field theory \ref\EY{T.~Eguchi and S.-K.~Yang, \mpl4 (1990)
1693.}.  The basic idea is first to modify the energy-momentum tensor so as to
obtain the one for the topological theory:
\eqn\Ttop{T_{\rm top}(z) ~=~ T(z) ~+~ \half \partial J(z) \ .}
The conformal weights of operators in the topological theory are thus given by
$h_{\rm top} = h_{N=2} - \half q$.  The supercurrents $G^+(z)$ and $G^-(z)$
have conformal weights $1$ and $2$ respectively in the topological theory.
Even though the conformal weights of these operators have changed I will still
use the mode labelling \SCAops\ of the $N=2$ superconformal theory.  To get the
physical spectrum of the topological theory one computes the cohomology of the
following (dimension zero) BRST charge:
\eqn\topbrst{Q ~=~ \oint G^+(z) ~dz \ .}
This charge is simply $G_{-\half}^+$; it manifestly satisfies  $Q^2 = 0$, and
it was shown in the last section that its cohomology can be represented by the
chiral primary fields. Thus the physical states of the topological theory are
precisely the chiral primaries, which now have (topological) conformal weight
equal to zero.  The reason why this twisted $N=2$ superconformal theory is
called topological is because the energy momentum tensor, $T_{\rm top}(z)$, is
BRST exact, {\it i.e.}
\eqn\Ttriv{T_{\rm top}(z) ~\equiv ~ \{Q \, , \, G^-(z)\} \ .}
Since $T_{\rm top}(z)$ (and $\widetilde T_{\rm top}(\bar z)$) generate
infinitessimal conformal transformations, including translations, it follows
that correlation functions of physical operators are all independent of their
locations.  Explicitly, if $\phi(z, \bar z)$ is a physical operator, then one
has:
$$
{\partial \over \partial z} ~ \phi(z, \bar z) ~=~ L_{-1} ~\phi(z, \bar z) ~=~
Q~(~G_{-\half}^- \phi) (z, \bar z) \ .
$$
In a correlation function of physical operators the contour integral \foot{ The
notation $\oint_z~~d \zeta$ means a small contour encircling a puncture at
$z$.} $Q = \oint_z G^+(\zeta ) ~d \zeta$ about $z$, can be deformed away from
$z$ to a contour encircling all the other punctures on the Riemann surface.
Since only physical operators are inserted at the punctures, and $Q$ kills all
of these operators, it follows that ${\partial \over \partial z} ~ \phi(z, \bar
z) \equiv 0$ is true as a Ward identity.  Similarly one also has ${\partial
\over \partial \bar z} ~ \phi(z, \bar z) \equiv 0$.

There is also another very important class of physical operators
\ref\DVV{R.~Dijkgraaf, E.~Verlinde and H.~Verlinde, \nup{352} (1991) 59.},
which is perhaps more accurately called a class of physical marginal
perturbations.  Let  $\phi(z, \bar z)$ be a chiral primary field.  Observe that
when $Q$, or its anti-holomorphic counterpart, $\widetilde Q$, acts upon
$(G_{-\half}^- \phi)(z, \bar z)$, $(\widetilde G_{-\half}^- \phi)(z, \bar z)$,
or $(G_{-\half}^- \widetilde G_{-\half}^- \phi)(z, \bar z)$, then the result is
a total derivative or zero.  These operators have topological conformal weights
$(h, \widetilde h)$ equal to $(1,0)$, $(0,1)$ and $(1,1)$ respectively.  As a
result, if $\Gamma$ is any closed curve, and $\Sigma$ is the Riemann surface,
then the integrals
\eqn\oneforms{\oint_\Gamma (G_{-\half}^- \phi)(z, \bar z)~dz\ , \qquad
\oint_\Gamma (\widetilde G_{-\half}^- \phi)(z, \bar z)~dz }
and
\eqn\margpert{\int_{\Sigma} d^2z ~ (G_{-\half}^-~\widetilde G_{-\half}^-
\phi)(z, \bar z) }
are physical operators.  Let $\phi_i (z, \bar z)$ be a basis for the chiral
priamary fields, and let $t_i$ be a set of parameters.  Define $\psi_i$ by:
\eqn\psidefn{\psi_i ~\equiv~ \big( G_{-\half}^-~\widetilde G_{-\half}^- \phi_i
\big) (z, \bar z) \ ,}
and introduce the perturbed topological corelation functions:
\eqn\topcor{ F_{i_1, \ldots, i_n} ~\equiv~ \bigg\langle~ \phi_{i_1}(z_1, \bar
z_1) \ldots \ldots \phi_{i_n} (z_n, \bar z_n)  ~  e^{-{\Big[ \sum_\ell t_\ell
\int\! d^2z\,  \psi_\ell (z, \bar z)}  \Big]} ~~\bigg\rangle \ .}
For the exactly the same reasons as outlined above, these correlation functions
are also independent of the locations of the insertion points, $z_1, \ldots,
z_n$.   As a result, the functions, $F_{i_1, \ldots, i_n}$, are totally
symmetric in their subscripts  $i_1, \ldots, i_n$.  However, these functions
depend upon the parameters, or `moduli', in a highly non-trivial manner.  The
properties of these functions have been extensively studied \DVV.  By suitably
differentiating the $F_{i_1, \ldots, i_n}$ one can generate correlators with
arbitrary insertions of the $\psi_j$.  To my knowledge there has been no
systematic study of correlators with insertions of the form \oneforms.  There
may well be some interesting topological interpretation for such correlators.
They should be related to the ``conformal blocks'' of the topological theory.
They might also lead to the braid matrices of the related non-topological
theories, but little has been done to develop these ideas.

\subsec{$N=2$ superconformal correlation functions and the topological model}

I now wish to relate the correlation functions \topcor\ to correlation
functions of chiral primary fields in the ``untwisted'' $N=2$ superconformal
model.   There are one or two minor subtleties that I wish to bring out into
the open. For simplicity I will restrict my attention to correlation functions
on the sphere.

First, one should note that the $U(1)$ current is anomalous in the topological
field theory:
\eqn\uoneanom{T_{\rm top}(z) ~ J(z)  ~=~ {-{c \over 6} \over (z-w)^3}  ~+~
{J(w) \over (z-w)^2} ~+~ {\partial J(z) \over (z-w)} \ . }
In terms of modes, one has
\eqn\anomcom{[L^{\rm top}_n\, ,\, J_m] ~=~ -m J_{m+n} ~-~ {c \over 6} ~n(n+1)
{}~\delta_{m+n, 0} \ .}
In particular, $J_0 = [J_1, L^{\rm top}_{-1}]$, but $J_0^\dagger =  - [J_{-1},
L^{\rm top}_{1}]$ $= J_0 + {c \over 3}$ .  Therefore, if $J_0 | 0 \!> = 0$ then
$0 = <\! 0 | J_0^\dagger $ $ = <\! 0 | (J_0 + {c \over 3})$.  Hence on the
sphere there is an anomaly of $-{c \over 3}$.  Consequently, for a topological
correlation function to be non-zero, the total $U(1)$ charge of all insertions
must be $+ {c \over 3}$.

In the original $N=2$ superconformal theory, the current $J(z)$ is not
anomalous and so to get corellators corresponding to \topcor\ one must
explicitly insert pure $U(1)$ fields whose total charge is $-{c \over 3}$.
There are, in principle, infinitely many ways to distribute this charge, but in
fact there are really only two natural methods.    The simplest way to
incorporate this negative charge is to insert the field $\bar \rho(\xi, \bar
\xi)$ at some point $\xi$.  Then one can show that:
\eqn\correl{\eqalign{\bigg\langle~ \phi_{i_1}(z_1, \bar z_1) \ldots\ldots
&\phi_{i_n} (z_n, \bar z_n) ~~\bar \rho(\xi, \bar \xi) ~ ~ e^{-\Big[ \sum_\ell
t_\ell \int\! d^2z\, |z ~-~ \xi|^{2(q_\ell - 1)} \psi_\ell (z, \bar z)  \Big]}
{}~~\bigg\rangle \cr ~ &=~ \left[ \prod_{p=1}^n~|z_{i_p} ~-~ \xi|^{-2q_{i_p}}
\right] ~  F_{i_1, \ldots, i_n} (t)  \ , }}
where $q_j$ is the $U(1)$ charge of $\phi_j$.  To see this one must suitably
modify the earlier argument since $Q$ doe not annihilate $\bar \rho(\xi, \bar
\xi)$.  Instead one uses:
\eqn\Qhat{ \widehat Q ~=~ \oint ~(\zeta - \xi)~ G^+(\zeta) ~d\zeta \ ,}
which now annihilates $\bar \rho(\xi, \bar \xi)$, but also generates some
algebraic mess:
\eqn\mess{\eqalign{\widehat Q~(~G_{-\half}^- \phi) (z, \bar z) ~& =~ \Big(
(z-\xi) ~ G_{-\half}^+ ~+~ G_{\half}^+ \Big) ~( G_{-\half}^- \phi) (z, \bar z)
\cr &=~ 2~\Big( (z-\xi) ~ L_{-1}  ~+~ (L_0 ~+~ \coeff12 J_0) \Big) ~\phi (z,
\bar z) \cr & =~ 2~(z - \xi)^{(1-q_i)}~\partial_z \Big[ (z-\xi)^{q_i} ~ \phi
(z, \bar z)  \Big] \ .}}
The measure in the integrals on the left hand side of \correl\ has therefore
been modified  so that when $\widehat Q$ hits an integrated operator, the
overall integrand is still a total derivative.   If one employs $\widehat Q$ in
the supersymmetry Ward identity argument one finds that $\partial_{z_j} \big[
{(z-\xi)}^{q_j} <~\ldots~> \big] = 0$.  From this one then arrives at \correl.

One should not be surprised at the factors of $(z-\xi)$ in \correl\ since in
the $N=2$ superconformal theory the operators $\phi_j$ have conformal weight
$h_j =\widetilde h_j = \half q_j $ while the operators $\psi_j$ have conformal
weight $h_j =\widetilde h_j = \half (1 + q_j)$.  Perturbing by such operators
breaks conformal invariance (for $q_j \not= 1$), and so the correlation
functions will care about the geometry of the surface, and in particular about
the location of $\xi$.

One should note that if the perturbations have $q_j =1$, then $\psi_j$ has
conformal weight $h_j =\widetilde h_j = 1$ in both the topological and the
untwisted $N=2$ superconformal theory.  This means that such operators provide
marginal perturbations perturbations of both theories, and do not violate
conformal invariance.  In these circumstances all the anomalous factors of
$(z-\xi)$ disappear from \correl.

Because of all the spurious factors of $(z-\xi)$, the left hand side of
\correl\ cannot, in general, be interpretted as a perturbed $N=2$
superconformal correlation function.  However, if one moves $\xi$ to $\infty$
on the complex plane then one can make the desired interpretation of \correl.
Recall that in order to get a proper finite limit, the correlation function
must  be multiplied by $|\xi|^{4h_\rho} = |\xi|^{2c \over 3} $ prior to sending
the operator $\bar \rho(\xi, \bar \xi)$ to infinity \ref\BPZ{A.A.~Belavin,
A.M.~Polyakov and A.B.~Zamolodchikov, \nup{241} (1984) 333.}.  By charge
conservation, one must have
$$
\sum_{p=1}^n ~q_{i_p} ~+~ \sum_{\rm perturbations} ~q(\psi_j) ~=~ {c \over 3 }
\ ,
$$
where the second sum is over the charges of all perturbations brought down in
the expansion of the exponential in \correl.  It is easily seen that the net
effect of rescaling the correlation function by $|\xi|^{2c \over 3} $ is
equivalent to replacing all factors of $|z - \xi|$ by $|{z \over \xi} - 1|$,
and as $\xi \to \infty$ these factors all go to unity.  Consequently, if  $\bar
\rho(\xi, \bar \xi)$ is sent to infinity on the complex plane, the mess
entirely disappears, and $F_{i_1, \ldots, i_n} (t)$ is precisely the perturbed
$N=2$ superconformal correlation function.  The coupling constants $t_j$ then
have a canonical, physical scaling dimension of $(1 - q_j)$.

An alternative identification with $N=2$ superconformal correlators can be
obtained by splitting the charge of $-{c \over 3}$ into two pieces.  That is,
one introduces the operator $\mu (z, \bar z) \equiv e^{-{i \over 2} \sqrt{c
\over 3} ~(X(z) ~+~ \widetilde X(\bar z))}$ at two distinct points $\xi_1$ and
$\xi_2$ on the sphere.  The operator $\mu$ in fact represents a particular
Ramond ground state.  Once again one generates many spurious factors of $(z -
\xi_m)$ in the correlation functions, but they can all be made to disappear by
conformally mapping to a flat cylinder  with $\xi_1$ and $\xi_2$ mapping to the
circles at either end of the cylinder.  Thus one also finds that $F_{i_1,
\ldots, i_n} (t)$ is exactly a perturbed $N=2$ superconformal correlation
function of chiral primary fields on the flat cylinder with Ramond ground
states on either end of the cylinder.  It is this interpretation that will be
the starting point of Vafa's lectures.

The basic point is that {\it topological correlation functions give you exact
perturbed $N=2$ superconformal correlation functions}, with the proviso that if
the perturbations break conformal invariance, you must choose an appropriate
two dimensional world sheet geometry.

Before leaving the subject of perturbed $N=2$ superconformal correlation
functions, I wish to make two important notes.  First, the perturbation:
$\sum_\ell t_\ell \int\! d^2z\,  \psi_\ell (z, \bar z) $ is not hermitian from
the point of view of the $N=2$ superconformal field theory.  The appropriate
hermitian perturbation is:
\eqn\hermpert{\Delta S ~ =~ -\sum_\ell \left \{ t_\ell \int\! d^2z\,  \big(
G_{-\half}^-~ \widetilde G_{-\half}^- \phi_\ell \big) (z, \bar z)  ~+~ \bar
t_\ell \int\! d^2z\,  \big( G_{-\half}^+~ \widetilde G_{-\half}^+ \bar
\phi_\ell \big) (z, \bar z) \right \} \ ,}
where $\bar t_\ell$ is the complex conjugate of $t_\ell$ and $\bar \phi_\ell$
is the anti-chiral field conjugate to $\phi_\ell$.  From the point of view of
topological field theory the second term in $\Delta S$ is BRST exact and so
decouples from all topological correletion functions.  Consequently, if one
made insertions of $e^{\Delta S}$ into all the correlations considered above,
all the terms involving $\bar t_\ell$ would vanish.  Thus $F_{i_1, \ldots, i_n}
(t)$ can still be interpretted as the appropriate chiral primary correlation
function in a perturbed superconformal theory with a perturbation of the form
\hermpert.  Thus we can use topological methods to compute correlation
functions in the massive quantum field theory whose action can be thought of as
$S_0 + \Delta S$, where $S_0$ is the formal  action of the original $N=2$
superconformal model \foot{ Many conformal field theories do not appear to be
derived from an action.  One should thus interpret this as a statement of how
to define the theory via perturbation theory.}.   Since the perturbing
operators in \hermpert\ are top components of superfields, it follows from the
general theory of supersymmetry that these massive quantum field theories are
still $N=2$ supersymmetric.

\nref\witjones{E.~Witten, \cmp{121} (1989) 351.}
\nref\DKL{L.J.~Dixon, V.~Kaplunovsky and J.~Louis, \nup{329} (1990) 27.}
\nref\VPAS{V.~Periwal and A.~Strominger, \plt{235B} (1990) 261; A.~Strominger,
\cmp{133} (1990) 163.}
\nref\CecVaf{S.~Cecotti and C.~Vafa, \nup{367} (1991) 359; \prl{68} (1992) 903;
\mpl{7} (1992) 1715.}
There is a minor cautionary note to be sounded at this juncture.  The
decoupling of the BRST trivial states is a somewhat subtle business.  One may
need to perform some mild regularization, and this may introduce contact terms.
 Because of rather general arguments for topological field theories, one knows
that it is possible to regularize theory so that BRST trivial states properly
decouple \wittop\witjones.  Equivalently, one can always find appropriate
representatives of the BRST cohomology and corrections to the BRST charge so
that all the correlators that should be zero {\it are} indeed zero.  The fact
that such choices are being made has been hidden in the discussion so far.  In
the $N=2$ superconformal theory the choices of representatives and the explicit
understanding of the regularization procedures and contact terms can lead to
extremely important insights into the geometry of the moduli space of the
theory (see, for example, \refs{\CGPA,\CGPB,\DKL{--}\CecVaf}).  Some of these
issues will be addressed indirectly in the next sections, and will also be
discussed by Vafa.

\nref\Loss{A.~Lossev, {\it ``Descendants constructed from matter field and K.
Saito higher residue pairing in Landau-Ginzburg theories coupled to topological
gravity,''} preprint TPI-MINN-92-40-T.}
\nref\Keke{K.\ Li, \nup{354} (1991) 711.}
\nref\VVtopgr{E.\ Verlinde and H.\ Verlinde, \nup{348} (1991) 457.}
\nref\DVVB{ R.~Dijkgraaf, E.~Verlinde and H.~Verlinde, {\it ``Notes on
Topological String Theory and 2-D Quantum Gravity,''}  Lectures given at Spring
School on Strings and Quantum Gravity, Trieste, Italy, Apr 24 -- May 2, 1990
and at Cargese Workshop on Random Surfaces, Quantum Gravity and Strings,
Cargese, France, May 28 -- June 1, 1990.}
\nref\Dubr{B.A.~Dubrovin, \cmp{145} (1992) 195;  {\it ``Integable systems and a
classification of two-dimensional topological field theories,''} preprint
SISSA-162-92-FM (1992), bulletin board hep-th@xxx.lanl.gov - 9209040; \nup{379}
(1992) 627. }
\nref\Kri{I.~Krichever, \cmp{143} (1992) 415;  {\it ``The tau function of the
universal Whitham hierarchy, matrix models and topological field theories,''}
preprint LPTENS-92-18 (1992).}
It is also interesting to point out that the required regularization involves
the metric on the underlying Riemann surface, and one cannot choose the
regulator uniformly over the entire moduli space of a Riemann surface.  As a
result, when one couples the foregoing topological matter models ({\it i.e} the
topologically twisted $N=2$ superconformal models) to topological gravity there
will be contact terms coming from the gravitational sector.  This important
observation \Loss\ appears to lead to a derivation of the coupling of
topological matter to topological gravity \refs{\Keke{--}\DVVB} directly from a
\LG formulation.   This may also be related to, and perhaps provide some
explanation of, the recent results of Dubrovin and Krichever  \refs{\Dubr,\Kri}
on the \LG derivation of the integrable hierarchies associated with topological
matter coupled to topological gravity.

\newsec{Properties of topological correlation functions}

It is the purpose of this section, and  to some extent of the the two following
sections, not only to describe the properties of topological correlation
functions, but also to show how the topological correlation functions can be
computed from rather limited knowledge of the underlying $N=2$ superconformal
theory, or its topologically twisted counterpart.

\subsec{General structure}

I will begin by simply enumerating and describing a number of properties of the
correlation functions \topcor.  My discussion here will essentially be a
summary of some of the results of \refs{\DVV, \DVVB}.  I will also only
consider correlation functions on the sphere.

Define two special correlation functions:
\eqn\metcijk{ \eta_{ij} ~\equiv F_{ij}(t) \ , \qquad C_{ijk} ~\equiv F_{ijk}(t)
\ . }
Since $F_{i_1, \ldots, i_n}$ is totally symmetric in all its indices, these
functions must also be totally symmetric.  For $t=0$ the function $\eta_{ij}$
forms an invertible topological metric.  To see this, suppose that we normalize
the chiral primary fields $\phi_i$ so that in the $N=2$ superconformal theory
one has:
$$
\big\langle \phi_i(z, \bar z) ~ \bar \phi_j(w, \bar w) \big\rangle ~=~
{\delta_{ij} \over |z-w|^{2h_i}} \ ,
$$
where $\bar \phi_j$ is the anti-chiral conjugate of $\phi_j$.  Now recall that
the spectral flow using $\bar \rho$ maps $\cR$ isomorphically onto $\bar \cR$,
and hence there is an invertible matrix ${V_i}^j$ such that:
$$
\bar \phi_j (w, \bar w) ~=~ {V_j}^k \lim_{z \to w} |z-w|^{4h_i} \bar \rho(z,
\bar z)  \phi_k (w, \bar w) \ .
$$
It then follows that the $N=2$ superconformal correlation function $< \phi_i
\phi_j \bar \rho>$ is proportional to the invertible matrix ${V_i}^j$, and
therefore one has $\eta_{ij} = {V_i}^j$.

I will use the metric $\eta_{ij}$, and its inverse $\eta^{ij}$, to raise and
lower indices.  One should note that ${C_{ij}}^k$ are precisely the structure
constants of the chiral ring.  To see this, simply take the limit $z_1 \to z_2$
in the three point function (remember that the correlation function is
independent of the $z_i$) in which case the three point function collapses to
the structure constants times the metric.

One can prove that the topological correlation functions, at general values of
$t$, have the following properties:

\item{(i)} $\eta_{ij}$ is in fact independent of $t$.
\item{(ii)} $F_{i_1, \ldots, i_n} ~=~ {C_{i_1 i_2}}^j~ F_{j, i_3, \ldots, i_n}$
\item{(iii)}$ {C_{i j}}^m ~ {C_{k l m}} ~=~ {C_{(i j}}^m ~ {C_{k l) m}}$
\item{(iv)} $\partial_\ell ~ C_{ijk} ~ =~ \partial_{( \ell} ~ C_{ijk)}$

\noindent where $( \ \ )$ denotes symmetrization of the indices enclosed, and
$\partial_\ell = {\partial \over \partial t_\ell}$.

In proving these properties, I will work with the $N=2$ superconformal
correlators, and take \correl\ as the definition of $F_{i_1, \ldots, i_n}$.
The fact that $\eta_{ij}$ is independent of $t$ can be demonstrated using the
Ward identities of $G^-(z)$.  One considers an insertion of $\psi_\ell = \big(
G_{-\half}^-~ \widetilde G_{-\half}^- \phi_\ell \big) (z, \bar z)$ in the
correlator defining $F_{ij}(t)$.  By writing this insertion in terms of a
contour integral $\oint_z {\cal V}(\zeta)  \ G^-(\zeta) \, d\zeta$ for a
suitable choice\foot{A suitable choice is to take $ {\cal V}(\zeta)= {{(\zeta -
z_1)(\zeta - z_2)} \over (\zeta - \xi)}$.  Since $G^-(\zeta)$ has conformal
weight $3/2$, the vector field $ {\cal V}(\zeta)$ can diverge, at most,
linearly as $\zeta \to \infty$ if the contour integral is to leave no residue
at infinity \BPZ.} of vector field, ${\cal V}(\zeta)$, one can pull the contour
off at infinity and arrange that it also annihilates $\phi_i$, $\phi_j$ as well
as other insertions of the perturbation, $\psi_k$.  One also needs to make use
of the fact that $G_{r}^- \bar \rho = 0$ for $r \ge -{3 \over 2}$ (see equation
\Grho).  Thus one can establish that all the {\it integrands} defining the
perturbations of $F_{ij}$ vanish, and so the metric $\eta_{ij}$ is a constant,
and hence flat, metric.

To establish (ii) one inserts a complete set of states on a circle that
separates $\phi_{i_1}$ and $\phi_{i_2}$ from the other fields $\phi_{i_\ell}$.
One then uses the Hodge decomposition theorem on this complete set of states.
The terms that appear in the completeness sum are either of the form a) $|\phi
\!><\! \phi|$, b) $G_{-\half}^+|\chi_1\!><\! \chi_1| G_{\half}^-$ or c)
$G_{\half}^-|\chi_2 \!><\! \chi_2| G_{-\half}^+$: where $|\phi \!>$ is chiral
primary.  Since  $G_{-\half}^+ = Q$, the insertion of anything of the form b)
or c) vanishes.  This leaves a sum only over the chiral primaries, and hence
estalblishes (ii).   The result (iii) then follows from considering all
possible ways of factorizing a four-point function.

It is straightforward, but technical, to prove the identity (iv).  It basically
follows from an elementary fact about conformal field theory: Any correlator
depends upon any four of the insertion points via the cross-ratio of those
points \BPZ.  As a consequence, if three points are fixed and the rest are
integrated over, one can transform one of the integrations to an integration
over the cross ratio, and thence transform it to an integration over one of the
previously fixed points.  Put another way, conformal invariance means that in
the perturbed three-point function with $n$ integrated insertions, one can
integrate over {\it any} subset of $n$ of the $n+3$ insertions.  There are then
two other elements that needed to complete the proof:  First, one must move the
operators $G_{-\half}^-$ and $\widetilde G_{-\half}^-$ from the old integrated
insertion to the new integrated insertion.  This is once again accomplished by
contour integration tricks as outlined above.  Secondly, in moving around these
operators and changing the integration variable one generates lots of factors
of $(z -\xi)$.  These all conspire to transform all the factors of $(z -\xi)$
in \correl\ in precisely the correct manner.

{}From (iv) it follows that one can write
\eqn\prep{C_{ijk} ~=~ \partial_i \partial_j \partial_k~{\cal F} \ ,}
for some analytic function ${\cal F}(t)$.  This function is called the {\it
free energy} of the model and completely characterizes all of the topological
correlation functions.  By scaling the fields and the coupling constants in the
three-point function one can show that ${\cal F}(t)$ must satisfy:
\eqn\Fscale{{\cal F}(\lambda^{(1 - q_j)}~t_j) ~=~ \lambda^{3 - {c \over 3}}
{\cal F}(t_j) \ ,}
where $q_j$ is the charge of the field $\phi_j$.

\nref\EVNW{E.~Verlinde and N.P.~Warner, \plt{269B} (1991) 96.}
\nref\Maas{Z.~Maassarani, \plt{273B} (1991) 457.}
It turns out that \prep, \Fscale\ and property (iii) provide a highly
overdetermined system of equations that in practice appear to completely
determine the $C_{ijk}$, and hence ${\cal F}(t)$ up to quadratic, linear and
constant terms.  The only input necessary appears to be the number and
dimensions of the chiral primaries, and the unperturbed vanishing relations.
One can then make an ansatz for ${\cal F}$ and solve the equations
\refs{\DVV,\DVVB,\EVNW,\Maas}.  This is extremely laborious in practice, and
there are short cuts and far better methods, as I will describe.

\newsec{Effective \LG potentials}

\subsec{Structure and properties of the potentials}

If the original $N=2$ superconformal field theory had an effective \LG
potential, $W_0(x_a)$, it follows that the perturbed theory will also have an
effective \LG potential, $W(x_a;t_j)$.  This is because our perturbations
involve only the top components of the chiral primary superfields and therefore
preserve the supersymmetry.  The operators, $x_a$, are some generators of the
perturbed chiral ring, $\cR$.  Let $p_i(x_a)$ denote some polynomials in the
$x_a$ that form a basis for $\cR$.  The statement that one has an effective \LG
potential, $W$, means that the ring multiplication is defined by
\eqn\pertmult{p_i(x_a) ~p_j(x_a) ~=~ {f_{ij}}^k ~ p_k(x_a) \qquad {\rm mod}
\qquad \left\{ {\partial W \over \partial x_a} \right\} \ ,}
where ${f_{ij}}^k$ are structure constants computed by simply multiplying
polynomials modulo the ideal generated by the partial derivatives ${\partial W
\over \partial x_a} $.

There is a natural basis inherited from the confomal point: namely the
$\phi_i$, for which one has
\eqn\phibase{\phi_i ~ \phi_j ~=~ {C_{ij}}^k(t) ~ \phi_k \ ,}
where  ${C_{ij}}^k(t)$ are the structure constants obtained by conformal
perturbation theory.  One can, of course, use these structure constants to
write the basis, $\phi_i$, as polynomials in the generators, $x_a$.  Since the
structure constants are functions of $t$, the $\phi_i$ when considered as
polynomials in $x_a$, will also be functions of $t$.  That is, one has $\phi_i
= \phi_i(x_a; t_j)$.  For example, if one has $\phi_2 = \phi_1^2$ at $t =0$,
one might find that ${C_{11}}^0 = t$ and ${C_{11}}^2 = 1$ and hence one would
have $\phi_2 = \phi_1^2 + t$ $ = x_1^2 + t$.  The statement that there is an
effective \LG potential implies that these functions, $\phi_i(x_a; t_j)$, must
satisfy:
\eqn\phiring{\phi_i(x_a; t_\ell)~ \phi_j(x_a; t_\ell) ~=~ {C_{ij}}^k(t) ~
\phi_k(x_a; t_\ell) \qquad {\rm mod} \qquad \left\{ {\partial W \over \partial
x_a} \right\}  \ ,}
where one now simplifies the left hand side of this equation using polynomial
multipliction modulo the ideal generated by the partials ${\partial W \over
\partial x_a}$.  This imposes a vast set of constraints on $W(x_a;t_j)$ and the
$\phi_i(x_a; t_\ell)$.  Indeed, given ${C_{ij}}^k(t)$, one finds that the
functions $W(x_a;t_j)$ and $\phi_i(x_a; t_\ell)$ are greatly overdetermined.
There are, however, still more constraints.

Since the perturbation has the form \hermpert\ one can see that, at each point
in parameter space, under an infinitessimal change of parameters one has
$\delta W = - \sum_j \delta t_j \phi_j$, or
\eqn\phidefn{\phi_j(x_a; t_\ell) ~=~ - {\partial \over \partial t_j }
{}~W(x_a;t_\ell) \ .}
Given the form of the perturbation one might be tempted to conclude that $W$ is
linear in the $t_j$, and so $W(x_a;t_j) =  W_0 + t_j \phi_j$.  This is
incorrect.  One can see this on the computational level by merely observing
that the ${C_{ij}}^k(t)$ are non-trivial functions of $t$, and hence the
$\phi_i(x_a; t_\ell)$ must also be non-trivial functions of $t$.  Thus
\phidefn\ repesents a collection of differential equations\foot{It is, by
construction, physically obvious that these equations satisfy the requisite
integrability condition.  However, with some work, one can also prove it from
the definition of the $\phi_i(x_a; t_\ell)$. } for $W(x_a;t_j)$, which is a
non-trivial, non-linear function of the $t_j$'s.  On the physical level, the
non-linear dependence of $W$ upon $t_\ell$ is the result of contact terms
\Loss.  The beauty of the approach employed here is that all the consistency
conditions on ${C_{ij}}^k(t)$, $\phi_i(x_a; t_\ell)$  and $W(x_a;t_j)$
completely determine these contact terms and so one can avoid the subtleties of
such computations.

To determine $W(x_a;t_j)$ one can find the ${C_{ij}}^k(t)$ as described earlier
and then use the structure constants to find the $\phi_i(x_a; t_\ell)$ and then
solve \phidefn\ and \phiring.  It is also valuable to employ the scaling
behaviour of $W(x_a;t_j)$:
\eqn\Wscale{W(\lambda^{\omega_a}~x_a;\lambda^{1 - q_j} ~t_j) ~=~ \lambda ~
W(x_a;t_j)\  .}
In practice, it is usually simplest to solve everything at once.   That is,
make ans\"atze for $W$ and ${\cal F}$ that are consistent with \Wscale\ and
\Fscale\ and then write down all the constraints arising from the identities:
${C_{i j}}^m ~ {C_{k l m}} ~=~ {C_{(i j}}^m ~ {C_{k l) m}}$ and from the
equations \phiring\ and \phidefn.  The result is a highly overdetermined system
of equations for the unknown constants and functions in the ans\"atze.  This
system can be solved for simple models, but is completely unmanageable in
general.  As we will see in the next section, there are simpler ways to solve
the system.

\subsec{Relationship to topological \LG models}

Rather than constructing the \LG potential for a perturbed $N=2$ superconformal
theory, one can start with the \LG potential as the fundamental object and
obtain a topological field theory directly \ref\topLG{C.~Vafa, \mpl{6}  (1991)
337.}.  I will not review this approach in any detail here, but simply describe
how it connects with my discussion.  One parametrizes the effective \LG
potential, $W(x_a;t_j)$, in any manner one chooses but with the restriction
that the partial derivatives, \phidefn, define a basis for the chiral ring.  As
before, the ``maximal'' chiral primary field can be represented by the hessian:
\eqn\hessianb{H(x_a;t_j) ~=~ det \left( {\partial^2 W(x_a;t_j) \over \partial
x_b \partial x_c } \right) \ .}
The metric of the theory can then be defined by
\eqn\metgij{g_{ij} ~=~ {C_{ij}}^{H} \ ,}
where the structure constants are defined by polynomial multiplication as in
\phiring, and ${C_{ij}}^{H}$ denotes the coefficient of $H(x_a;t_j)$ in the
product of $\phi_i(x_a; t_\ell)$ and $\phi_j(x_a; t_\ell)$.  The fact that the
maximal element of the chiral ring is only defined up to an overall scaling
factor means that the metric \metgij\ will only be conformally related to the
natural topological metric, $\eta_{ij}$, introduced earlier. That is,
\eqn\conftrf{\eta_{ij} ~=~ \Omega(t_j)~g_{ij}\ , }
for some function, $\Omega(t_j)$, of the coupling constants.  One should also
note that because the coordinates $t_j$ are now arbitrary, the metric
$\eta_{ij}$, while (locally) flat, is not necessarily constant.  One can
determine the conformal factor $\Omega(t_j)$ by requiring that $\eta_{ij}$ be
flat, and one can reconstruct the ``flat coordinates'' of conformal
perturbation theory by solving the geodesic equations and constructing Gaussian
coordinates.

Thus in the topological \LG approach one, to some extent, loses sight of the
natural parametrization offered by conformal perturbation theory.  One must
reconstruct this parametrization by once again solving some rather unpleasant
equations.   One might of course wonder why one is so interested in this form
of the parametrization.  First, such coordinates are precisely those supplied
by conformal perturbation theory and therefore they are important in the
analysis of perturbed $N=2$ superconformal models.  Secondly, such coordinates
are central to the study of mirror symmetry and to the coupling of topological
matter models to topological gravity.

\subsec{Example:  minimal models}

For the minimal models, the basis for the {\it unperturbed} chiral ring (at
$t=0$) will be taken to be $\phi_\ell = x^\ell$, $\ell = 0,1, \ldots, n$.  For
convenience I will normalize $W_0(x)$ to
$$
W_0(x) ~=~ {1 \over (n+2)} ~ x^{n+2} \ .
$$
With this choice of basis one has
$$
{C_{ij}}^k(t = 0) ~=~ \delta_{i+j,k} \qquad 0 \le i,j,k \le n \ .
$$
To first order in perturbation parameters one has
\eqn\fstord{W ~= W_0 ~-~ A ~\left( \sum _{k=0}^n ~t_k~ x^k \right) \ ,}
where $A$ is an overall normalization of the perturbing field, and will be
fixed later.  The general perturbed superpotential, $W$, satisfies \Wscale\
with $\omega = 1/(n+2)$ and $q_j = j/(n+2)$.

The topological metric is given by $\eta_{ij} = <x^i x^j> = <x^{i+j}>$.
Normalize $x$ so that $<x^n> =1$, and then one has
$$
\eta_{ij} ~=~ \delta_{i+j,n} \ .
$$

To deduce ${C_{ij}}^k(t)$ and $W(x;t_\ell)$ from consistency conditions alone
is very painful, and so I will take a short-cut.  For the chiral primary field
$\phi_\ell$ consider

\eqn\nullst{X_p \equiv G_{-p-{1 \over 2}}^- ~ G_{-p+{1 \over 2}}^- ~ \ldots
G_{-{1 \over 2}}^- ~ \phi_\ell \ .}
This operator has conformal weight $h = \half (p + 1)^2 + \half {\ell \over
(n+2)}$ and $U(1)$ charge $q = {\ell \over (n+2)} - (p+1)$.  For $p \ge \ell$
this violates the unitarity bound\foot{ This unitarity bound follows from the
fact that the $L_0$ eigenvalue of a state must be at least that of its $U(1)$
component, {\it i.e} the energy-momentum tensor orthogonal to the $U(1)$
direction, $T(z) - (- \half (\partial X(z))^2)$, must also give rise to
non-negative conformal weights.}  $h \ge {3 \over 2c} q^2$ and so $X_\ell$ must
vanish identically.  Putting it another way, one can easy check that $X_\ell$
defines a null state in the $N=2$ superconformal minimal model.  Now consider a
 correlation function of the form
$$
\bigg\langle~ \phi_{i}(z_1) \phi_{j} (z_2)  \phi_{k} (z_3)~\left[ \prod_{i=1}^M
{}~ \Big(~G_{-\half}^- \phi_{i_m} \Big) (\zeta_m )\right]  ~\bar \rho(\xi)
{}~~\bigg\rangle \ .
$$
By the usual contour integration games and by clever choices of vector fields,
one can move all the $G^-$ operators  onto any one of the $\phi$'s.  In so
doing, the moding of the $G^-$ operators becomes exactly of the form \nullst.
As a result, this correlation function vanishes identically if $M$ exceeds the
minimum of $i,j,k$ and $i_1,i_2, \ldots, i_M$.  It follows immediately that
$C_{1ij}$ is at most linear in all the $t$'s.

By scaling one can see that $C_{1ij}$ must be proportional to $t_k$ with $k=
2n+1 - i - j$.  Fix the normalization constant $A$ in \fstord\ by taking
$C_{1nn} = t_1$.  Using property (iv) of the $C_{ijk}$ one then has $C_{11n} =
t_n$ and hence
\eqn\Cmatform{{C_{1n}}^0 ~=~ t_1 \ , \qquad {C_{1n}}^{n-1} ~=~ t_n \ .}
Now recall that at $t=0$ one has ${C_{ij}}^k = 1$ for $k = i+j$ and ${C_{ij}}^k
= 0 $ for $k > i+j$ and for $k= i+j -1$.  This remains unchanged for $t \not =
0$ since there is no $t_\ell$ with the requisite scaling dimension to modify
these particular structure constants.  Now use
\eqn\recdefC{{C_{1i}}^j ~ {C_{kj}}^\ell  ~=~ {C_{1k}}^j ~ {C_{ij}}^\ell }
to recursively determine the terms in ${C_{ij}}^k$ that are linear in $t_1$.
For example, take $i=1$ and use \recdefC\ to conclude that the matrix $C_2
\equiv {{(C_2)}_i}^j$ is of the form $(C_1)^2 + a t_n I$, where $I$ is the
identity and $a$ is some undetermined constant.  Continuing, one readily
establishes that
$$
{C_{j (n+1 - j - \ell)}}^\ell ~=~ t_1 \, \quad \ell = 0,1,\ldots, j-1 \ ,
$$
and using property (iv), it follows that
$$
{C_{1 (n+1 - j - \ell)}}^\ell ~=~ t_j \, \quad \ell = 0,1,\ldots, j-1 \ ,
$$
or as a matrix:
$$
C_1 ~=~ \pmatrix{
0&1&0&0&0&\ldots\ldots\ldots  &0 \cr
t_n&0&1&0 & 0&\ldots\ldots  \ldots &0 \cr
t_{n-1}&t_n&0&1&0&\ldots \ldots\ldots &0 \cr
\vdots&\vdots&\vdots&\vdots& \vdots & \ddots &\vdots\cr
\vdots&\vdots&\vdots&\vdots& \vdots & \ddots &0 \cr
\vdots&\vdots&\vdots&\vdots& \vdots & \ddots &1 \cr
t_1&t_2& t_3& \ldots &\ldots&\ldots \ldots\ldots  t_n & 0} \ .
$$
It is then elementary to determine $W(x;t_\ell)$ since ${dW \over dx}$ is the
polynomial of order $n+1$ in $x$ that must vanish in the chiral ring, but
${C_{1i}}^j$ are precisely the structure constants for multiplication by $x$,
and so ${dW \over dx}$ must be a multiple of the characteristic equation of the
matrix $C_1$.  Indeed, with my normalizations:
\eqn\Wprimedet{{dW \over dx} ~=~ det(x ~-~ C_1) \ .}
One can now easily construct the $\phi_j(x; t_\ell)$ recursively using the
structure of $C_1$.  One has $\phi_1 = x$ and:
\eqn\examphi{\eqalign{ x~\phi_j ~&=~ {C_{1j}}^k \phi_k ~=~ \phi_{j+1} ~+~
\sum_{\ell = 0}^j {C_{1j}}^\ell ~ \phi_\ell \cr ~&=~ \phi_{j+1} ~+~ \sum_{\ell
= 0}^j t_{n+1+\ell -j} ~ \phi_\ell \ , } }
which completely determines $\phi_{j+1}$.  This means that the effective
potential $W(x; t_\ell)$ can also be fixed using \phidefn.

In a later section I will need two special cases of this effective potential:
a) $t_1 = t$, $t_\ell = 0$ for $\ell \ge 2$; and  b) $t_n = t$, $t_\ell = 0$
for $\ell \le n-1$.  The first of these is completely trivial since $W(x;t)$
can be entirely determined by dimensional analysis:
\eqn\Wpota{ W(x;t) ~=~ {1 \over (n+2)} ~x^{n+2} ~-~ t x \ .}
The second is far less trivial.  One finds that $W(x;t)$ is a Chebyshev
polynomial.  Explicitly, one can write\foot{This formula can be established by
developing recursion relations for the determinants \Wprimedet.} :
\eqn\Wpotb{ W(x;t) ~=~ 2t^{{1 \over 2} (n+2)} ~ cos ( (n+2) \theta) \ ; \qquad
{\rm where} \quad x~=~ 2 \sqrt{t} ~cos\theta \ .}
The critical points of these potentials occur at $\theta = {j \pi \over n+2}$,
and at these points $W(x;t)$ takes the values $2 t^{{1 \over 2} (n+2)} cos(j
\pi)$ $= (-1)^j 2  t^{{1 \over 2} (n+2)} $.  This observation will be of
importance later.

\newsec{Flat Coordinates, Flat Bundles and Classical Integrable Hierarchies}

The purpose of this section is to review briefly some of the more recent
developments in the technology for computing the flat coordinatization of
effective potentials.  As this section is something of a digression from my
main objective, it may be ignored if the reader so desires.

\nref\Griff{P.\ Griffiths, {\it Ann. Math.} {\bf 90} (1969) 460.}
\nref\Saito{K.\ Saito, {\it Publ. RIMS} , Kyoto Univ., {\bf 19} (1983) 1231.}
\nref\Noumi{M.~Noumi, {\it Tokyo J. Math} {\bf 7} (1984) 1.}
\nref\BBAV{B.~Blok and  A.~Varchenko, \ijmp{7} (1992) 1467.}
\nref\ACSF{A.C.~Cadavid and S.~Ferrara, \plt{267B} (1991) 193.}
\nref\LSW{W.~Lerche, D.-J.~Smit and N.P.~Warner, \nup{372} (1992) 87.}
\nref\morrison{D.~Morrison, {\it ``Picard-Fuchs equations and mirror maps for
hypersurfaces''}, Duke preprint  DUK-M-91-14 (1991); hep-th@xxx.lanl.gov -
9111025.}
We have seen that, even in the simplest cases, it is a considerable labour to
compute the flat coordinatization of the effective potential.  There are,
however, more powerful techniques that can be borrowed from complex geometry
and from singularity theory.  These techniques are finding considerable use in
the analysis of mirror symmetry in Calabi-Yau manifolds and also in the
coupling of topological matter to topological gravity.  The basic idea probably
originated in Griffith's paper on complex surfaces in $\IC \IP^n$ \Griff, and
has since been entensively studied under the general headings of `variation of
Hodge structure' and `Gauss-Manin connections'.  The specific application to
singularity theory may be found in a nearly impenetrable paper by Saito \Saito\
and works by Noumi (see, for example, \Noumi).  There has also been
considerable discussion of the methods in the physics literature, initiated by
\BBAV\ and since developed in a number of places  \refs{\ACSF{--}\morrison}.

Once again, let  $W(x_a;t_j)$ be a general perturbation of $W_0(x_a)$, and
define $\phi_i(x_a;t_\ell)$ by \phidefn.  Consider $x_a$, $a =1,2,\ldots, N$,
to be complex variables and introduce ``formal'' integrals of the form
\eqn\period{u_i^{(\lambda)} ~\equiv~ (-1)^{(\lambda+1)} ~\Gamma(\lambda+1)
\int_X {\phi_i(x_a;t_\ell) \over W(x_a;t_\ell)^{\lambda+1}} \ .}
In this expression, $\lambda$ is a complex parameter and $\Gamma$ is the usual
gamma function.  The prefactor has been introduced for convenience to soak up
irritating factors arising from integration by parts.  The surface, $X$, over
which the integral is to be performed is any  real $(2N-1)$-dimensional
hypersurface (cycle) around some component of the surface defined by $W=0$.
The only thing that one really needs to know about these integrals is that the
integral of any total derivative is zero.

Differentiating $u_i^{(\lambda)}$ with respect to $t_j$ one obtains two terms:
$$
\eqalign{ {\partial \over \partial t_j}~u_i^{(\lambda)} ~=~ &- (-1)^{\lambda+2}
{}~\Gamma(\lambda+2) \int_X { {\phi_i(x_a;t_\ell) ~\phi_j(x_a;t_\ell)} \over
W(x_a;t_\ell)^{\lambda+2}} \cr &~+~ (-1)^{\lambda+1} ~\Gamma(\lambda+1) \int_X
{\partial_j \phi_i(x_a;t_\ell) \over W(x_a;t_\ell)^{\lambda+1}} \ .}
$$
One now decomposes $\phi_i(x_a;t_\ell) \phi_j(x_a;t_\ell)$ modulo $\cR$, {\it
i.e.}:
$$
\phi_i(x_a;t_\ell) ~ \phi_j(x_a;t_\ell) ~=~ {C_{ij}}^k ~\phi_k(x_a;t_\ell) ~+~
q_a~{\partial W \over \partial x_a} \ ,
$$
where $q_a$ are some polynomials.  Integrating by parts one obtains:
$$
{\partial \over \partial t_j}~u_i^{(\lambda)} ~=~ - {C_{ij}}^k ~
u_k^{(\lambda+1)}  ~+~ (-1)^{\lambda+1} ~\Gamma(\lambda+1) \int_X { \partial_j
\phi_i(x_a;t_\ell) + {\partial \over \partial x_a} q_a   \over
W(x_a;t_\ell)^{\lambda+1}} \ .
$$
One may have to reduce the numerator of the second term modulo $\cR$, and once
again integrate by parts.  Repeating this process as often as is necessary, one
obtains an equation of the form:
$$
{\partial \over \partial t_j}~u_i^{(\lambda)}  + {C_{ij}}^k ~ u_k^{(\lambda+1)}
- \sum_{n=0}^\infty {\Gamma_{(n)}}^k_{ij} ~ u_k^{(\lambda - n)} ~=~ 0\ .
$$
In other words the $u_i^{(\lambda)}$ are flat sections of a trivial bundle over
the parameter space of deformations of the superpotential $W$. The foregoing
equation defines the Gauss-Manin connection on this space.  One can write down
the integrability condition of the connection and break it into components
corresponding to the superscript $(\mu)$ on $u_i^{(\mu)}$.  The resulting
flatness equations require the ${C_{ij}}^k$ to satisfy property (iii) and the
covariant analogues of property (iv). (The structure constants, ${C_{ij}}^k$,
trivially satisfy property (iii) since the structure constants were defined
here using polynomial multiplication.) The affine connection in these covariant
derivatives is given by $\Gamma^k_{ij} = {\Gamma_{(0)}}^k_{ij}$.   The fact
that the system of equations that we wish to solve comes from the integrability
conditions of a linear system, means that we are dealing with some form of
classical integrable hierarchy.

At this point, there is something of a technical minefield, which can probably
be passed so as to obtain a general result.  However, the following partial
results are know.  If one restricts to relevant perturbations (so that the
coupling constants have strictly positive scaling dimensions) then the flat
coordinates we seek are precisely those obtained by requiring that the
connection vanishes.  In other words, parametrize the potential with arbitrary
functions of {\it flat} coordinates, and then set $\Gamma^k_{ij}$ to zero.  The
result is a system of differential equations that define the arbitrary
functions in terms of flat coordinates.  One can also incorporate marginal
parameters ({\it i.e.} ones with vanishing scaling dimension) into this
procedure.  One modifies the starting point so as to incorporate an analogue
the conformal factor in \conftrf.  Specifically, one starts with
\eqn\unought{u_0^{(\lambda - 1)} ~\equiv~ (-1)^\lambda~\Gamma(\lambda) \int_X
{\omega(t_\ell) \over W(x_a;t_\ell)^{\lambda}} \ ,}
where $\omega(t_\ell)$ is a function {\it only } of the marginal (dimension
zero) parameters.  One then defines $u_i^{(\lambda)} ~=~ \partial_i
u_0^{(\lambda - 1)}$, and proceeds as above.  One then sets {\it all} of the
connection terms, ${\Gamma_{(n)}}^k_{ij}$, to zero, and as well as determining
the flat coordinates, this also determines the function $\omega(t_\ell)$.  The
problem arises if one tries to incorporate irrelevant perturbations (with
parameters of negative scaling dimension).  It is not yet known (at least to
physicists) how to deal with these.  In particular, the flat coordinate
equations cannot easily be separated out of the foregoing procedure.  If one
reads \Saito\ one is left with the impression that there should not be a
problem, but so far I am not aware of a successful implemation the abstractions
of \Saito\ into a computable procedure.

The foregoing procedures for computing flat coordinates are described in
considerable detail in \LSW.  It turns out that this method is by far the most
efficient for computing flat coordinates, and to illustrate this I will show
how it yields a different formulation of the flat coordinates for the minimal
models.

One starts from
\eqn\uzero{u_0^{(\lambda - 1)} ~\equiv~ (-1)^{(\lambda)} ~\Gamma(\lambda)
\int_X {1 \over W(x;t_\ell)^{\lambda}} \ ,}
where the integration is taken over any contour about one or more zeros of $W$.
Differentiating, one obtains:
$$
\eqalign{ {\partial^2 \over \partial t_i \partial t_j} u_0^{(\lambda - 1)} ~=&~
(-1)^{(\lambda+1)} ~\Gamma(\lambda+1) \int_X {(\partial_i \partial_j W) \over
W(x;t_\ell)^{\lambda+1}} \cr   &~+~  (-1)^{(\lambda+2)} ~\Gamma(\lambda+2)
\int_X {(\partial_i W)~(\partial_j W) \over W(x;t_\ell)^{\lambda+2}} \ .}
$$
One must now rewrite the second numerator as follows:
\eqn\pijdefn{  (\partial_i W)~(\partial_j W) ~=~ {C_{ij}}^k ~ (\partial_k W)
{}~+~ p_{ij} \partial_x W \ . }
Integrating by parts, the connection term is given by:
\eqn\connec{(-1)^{(\lambda+1)} ~ \Gamma(\lambda+1) \int_X {(\partial_i
\partial_j W)  - \partial_x p_{ij} \over W(x;t_\ell)^{\lambda+1}} \ .}
Since $\partial_i W$ is a polynomial of degree at most $n$, it follows that
$p_{ij}$ has degree at most $n-1$, and $\partial_i \partial_j W$ has degree at
most $n-2$.  Consequently the numerator in \connec\ is of degree at most $n-2$,
and so can be written in terms of elements of $\cR$.  Thus there are no further
integrations by parts that need to be done. Flat coordinates are then defined
by imposing:
$$
\partial_i \partial_j W  ~=~ \partial_x p_{ij} \ .
$$
It is now convenient to consider formal power (and Laurent) series in the
variable $x$, and to introduce the notation $[\ \ ]_+$ to mean that one should
discard all the negative powers of $x$ that appear in the expansion of the
quantity in the square brackets.  In particular, $W^{j \over n+2}$ is to be
thought of as a formal expansion in decreasing powers of $x$ and starting with
$x^j$.  From \pijdefn\ it follows that the $p_{ij}$ can be written as:
$$
p_{ij} ~=~ \left[ {(\partial_i W)~(\partial_j W) \over (\partial_x W)}
\right]_+ \ ,
$$
and hence the coefficients of $W$ must satisfy the differential equation
\eqn\deqa{\partial_i \partial_j W ~=~ \partial_x \left[ {(\partial_i
W)~(\partial_j W) \over (\partial_x W)} \right]_+ \ .}
This equation can be greatly simplified using the scaling property \Wscale\ of
$W$, which implies:
\eqn\scalea{ x~{\partial W \over \partial x} ~+~ \sum_{j=0}^{n} (n + 2 -j)~t_j~
{\partial W \over \partial t_j} ~=~ (n+2)~W \ ,}
and
\eqn\scaleb{ x~\partial_x \left({\partial W \over \partial t_j} \right)  ~+~
\sum_{i=0}^{n} (n + 2 -j)~t_i~{\partial^2 W \over \partial t_i \partial t_j}
{}~=~ j~{\partial W \over \partial t_j} \ .}
Multiplying both sides of \deqa\ by $(n + 2 -j)~t_i$ and summing, and then
simplifying a little, one obtains:
$$
\left({j+1 \over n+2} \right) ~ {\partial W \over \partial t_j}  ~=~ \partial_x
 \left[ {W~(\partial_j W) \over (\partial_x W)} \right]_+ \ .
$$
This can be rearranged to give:
$$
\left[  \left({j+1 \over n+2} ~-~ 1 \right) ~ {\partial W \over \partial t_j}
{}~+~ {W (\partial_x^2 W)~(\partial_j W) \over (\partial_x W)^2} ~-~ { W
(\partial_x \partial_j W) \over (\partial_x W)} \right]_+ ~=~ 0 \ .
$$
If one ignores the $[\ \ ]_+$, then one can arrange this last equation into a
collection of logarithmic derivatives and conclude that $\partial_j W = A
\partial_x W^{j+1 \over n+2}$ for some constant $A$.  If one is careful about
the $[\ \ ]_+$ then one simply obtains the equation
\eqn\DVVeqn{\partial_j W ~=~ A ~\partial_x \left[ W^{j+1 \over n+2} \right ]_+
\ ,}
where $A$ is a constant.  This is precisely the differential equation whose
solution and properties can be connected directly with the KdV hierarchy of the
matrix model \DVV.

\newsec{$N=2$ Supersymmetric quantum integrable models}
\nref\BG{A.~Bilal and J.-L.~Gervais, \plt{206B} (1988) 412; \nup{314} (1989)
646; \nup{318} (1989) 579; \nup{326} (1989) 222.}
\nref\Bilal{ A.\ Bilal, \nup{330} (1990) 399;  \ijmp{5} (1990) 1881.}
\nref\HM{T.~Hollowood and P.~Mansfield, \plt{226B} (1989) 73; \nup{330} (1990)
720.}
\nref\EY{T.~Eguchi and S.-K.~Yang, \plt{224B} (1989) 373; \plt{235B} (1990)
282.}
\nref\FLMW{P.~Fendley, W.~Lerche, S.D.~Mathur and N.P.~Warner, \nup{348} (1991)
66.}
\nref\WLNW{W.~Lerche and N.P.~Warner, \nup{358} (1991) 571.}
\nref\NemWar{D.~Nemeschansky and N.P.~Warner, \nup{380} (1992) 241.}
\nref\PFKI{P.~Fendley and K.~Intriligator, \nup{372} (1992) 533; \nup{380}
(1992) 265.}

It is known that many of the $N=2$ superconformal models have one (and
frequently more than one) perturbation that leads to an $N=2$ supersymmetric
quantum integrable field theory.  There are several methods for seeing this,
for example,   one can use conformal perturbation theory \ref\Zama{A.B.
Zamolodchikov, {\it JETP Letters}  {\bf 46} (1987) 161; {\it ``Integrable field
theory from conformal field theory,''} in  Proceedings of the Taniguchi
symposium (Kyoto 1989), to appear in Adv. Studies in Pure Math; R.A.L. preprint
89-001; \ijmp{4} (1989) 4235.} as Mussardo described in his lectures, but this
approach is unsystematic since one often does not know which perturbation to
consider, and what spin or form the non-trivial conserved currents will have.
On the other hand, there are more sophisticated Toda and free field methods
that lead to families of integrable models.  There is now a vast literature on
this subject (see, for example, \refs{\BG {--} \EY}).  In particular, much is
now known about the perturbations of $N=2$ superconformal coset models that
lead to quantum integrable field theories (see, for example, \refs{\FMVW, \FLMW
{--} \PFKI}).  The basic rule of thumb is that if there is some form of
$W$-algebra, and if the $W$-algebra generators are the top components of a
superfield, then there is usually special $N=2$ supersymmetry preserving,
relevant perturbation that leads to an $N=2$ supersymmetric integrable model.

In the minimal series there are three known perturbations that separately lead
to an integrable model.  In the notation of section 5, these perturbations are
given by $\phi_1$, $\phi_2$ and $\phi_n$.  For the $\phi_1$, or most relevant,
perturbation, the states corresponding to the first few non-trivial conserved
currents are:
$$
\eqalign{& G_{-\half}^+~G_{-\half}^- ~ J_{-1}^2 ~| 0 \! > \ , \cr &
G_{-\half}^+ ~G_{-\half}^-[ ~J_{-1}^3 ~+~ \coeff{1}{2} (c-3)  ~ J_{-1} ~L_{-2}]
{}~| 0 \! > \ , \cr & G_{-\half}^+ ~G_{-\half}^-[ ~J_{-1}^4 ~+~  (c-3) L_{-2}
{}~ J_{-1}^2 ~+~\coeff{2}{9} (c^2 - 3c +18) ~L_{-2}^2 ]~ | 0 \! > \ ;}
$$
where $c = {3n \over n+2}$ is the central charge of the model.  Note that all
these states are top components of superfields.  (They are also $W$-generators
in the sense that the $N=2$ superconformal minimal models have a $W$-algebra
embedded in them, but the all the $W$-generators can be written in terms of
$G^\pm (z)$, $J(z)$ and $T(z)$.)

\nref\ALDNNW{A.~LeClair, D.~Nemeschansky and  N.P.~Warner, USC preprint
USC-92/010, Cornell preprint CLNS 92/1148, to appear in Nucl.~Phys ~B.}
Once one has a quantum integrable model the basic objects of physical interest
are the soliton spectrum and the corresponding scattering matrix.  To determine
these is often something of an art-form, and this problem will be discussed in
other lectures at this school.  My purpose here is to show how, in $N=2$
supersymmetric theories, the \LG description yields much more complete
information about the soliton spectrum.  It is then possible to use this
information to compute more easily the soliton scattering matrices (see
\refs{\PFKI,\ALDNNW} and Nemeschansky's lectures at this school).

\subsec{ Bogolmolny bounds for kinks}
\nref\WLNWB{W.~Lerche and N.P. ~Warner, {\it ``Solitons in integrable, $N=2$
supersymmetric \LG models,''} in {\it Strings and Symmetries, 1991}, editors:
N.~Berkovits, H.~Itoyama {\it et. al.}, World Scientific, 1992.}

For the moment I will not assume that the model is integrable, all I will
assume is that the model is obtained by some relevant perturbation of an $N=2$
superconformal theory, and that the ground states of the theory are all fully
resolved ({\it i.e.} mathematically non-degenerate) \foot{That is, all the
small oscillations about these ground states are massive.}.  As was observed
earlier, since the bosonic potential of the theory is given by $V= |\nabla
W|^2$, the ground states of the theory all have zero-energy and occur at the
critical points of $W$.  Let $x_a^{(\alpha)}$ denote the values of the \LG
fields  at these critical points, and suppose that $x_a^{(\alpha)}$ and
$x_a^{(\beta)}$ are two distinct such points.  The solitons are found by
seeking the minimum energy ``kinks'' that interpolate between the two
correponding \LG vacuum states.  That is, I now consider a Lorentzian,
two-dimensional field theory on $\IR^2$, with a spatial coordinate, $\sigma$,
and time coordinate, $\tau$, and seek classical configurations that have:
\eqn\bndcond{ x_a ~=~ x_a^{(\alpha)}\quad {\rm at} \quad \sigma ~=~ - \infty
\quad\ \  {\rm and}\ \ \quad x_a ~=~ x_a^{(\beta)}\quad {\rm at} \quad \sigma
{}~=~ + \infty  \ .}
One has no idea of what the kinetic term is for the model in question since the
kinetic term does renormalize.  However, quite independent of this kinetic term
one can still deduce some properties of the soliton spectrum.  First, one can
obtain a Bogomolny bound that states that any configuration that satisfies the
boundary conditions \bndcond\ must have a mass $m_{\alpha \, \beta}$ that obeys
the bound
\eqn\bogbnd{ m_{\alpha \, \beta} ~\ge~ |\Delta W | \quad {\rm where} \quad
\Delta W ~\equiv~ W(x_a^{(\beta)}) ~-~ W(x_a^{(\alpha)}) \ .}
Moreover, one has $m ~=~ |\Delta W |$ if and only if the soliton is a {\it
fundamental soliton} in that it is annihilated by two of the four supercharges
of the perturbed theory.  Furthermore, the classical trajectory,
$x_a^{(\alpha)}(\sigma)$, of a fundamental soliton at rest, maps to a straight
line in the complex $W$-plane, {\it i.e.} the complex phase of $W(x_a
(\sigma))$ is fixed for all values of $\sigma$.  A semi-classical proof of
these statements can be found in \FMVW, and they are an elementary
generalization of arguments given in \ref\OlWit{D.~Olive and E.~ Witten.
\plt{78} (1978) 97. }.  Here I will outline the exact quantum proof of the
Bogolmolny bound \refs{\WLNW,\WLNWB}.

One can show that in the presence of the perturbation \hermpert, the
supercurrents receive corrections so that the corresponding conservation laws
become:
\eqn\pertsuper{\eqalign{
\del_{\bar z} G^+(z,\bar z)\ &=\ \sum_i ~ t_i\ (1 - q_i)\ \del_z
\big(\widetilde G^-_{-{1 \over 2}}\phi_i \big)(z,\bar z)\cr
\del_{\bar z} G^-(z,\bar z)\ &=\ \sum_i ~ t_i\ (1 - q_i)\ \del_z
\big(\widetilde G^+_{-{1 \over 2}}\bar \phi_i \big)(z,\bar z)\cr
\del_z \widetilde G^+(z,\bar z)\ &=\ \sum_i ~ t_i\ (1 - q_i)\ \del_{\bar z}
\big(G^-_{-{1 \over 2}}\phi_i \big)(z,\bar z) \cr
\del_z \widetilde G^-(z,\bar z)\ &=\ \sum_i ~ t_i\ (1 - q_i)\ \del_{\bar z}
\big(G^-_{-{1 \over 2}}\bar \phi_i \big)(z,\bar z) \cr} }
This can be established using perturbation theory, and it can also probably be
established using the general properties of $N=2$ supersymmetry.  The usual
arguments of conformal perturbation theory (see Mussardo's lectures or see
\Zama) are only easily implemented to first order in the perturbation.  For
many purposes this is usually enough, but it is insufficient to establish
\pertsuper\ in all generality at all points in $t$-parameter space.  However,
one can generalize the arguments of \Zama\ to a general point in $t$-parameter
space, and having done this, first order perturbation theory is sufficient.
The desired result then follows from this first order perturbation theory and
the observation that $\delta W = \phi_i \delta t_i$ at all points in
$t$-parameter space.

In the perturbed theory one now has four conserved charges, $Q_+$, $Q_-$,
$\widetilde Q_+$ and $\widetilde Q_-$:
\eqn\consG{\eqalign{Q_+ \ =&\ \int G^+\ dz ~-~ \int ~ \sum_i ~t_i ~(1 - q_i ) \
(\widetilde G^-_{-{1 \over 2}}) \phi_i (z, \bar z) \ d\bar z \cr
\widetilde Q_+ \ = &\ \int \widetilde G^+\ d\bar z ~-~ \int ~ \sum_i ~t_i ~(1 -
q_i ) \ (G^-_{-{1 \over 2}} \phi_i) (z, \bar z) \ d z  \ , }}
and similarly for $Q_-$ and $\widetilde Q_-$.  The general structure of such an
$N=2$ supersymmetry algebra is:
\eqn\comalg{\eqalign{
\big\{Q_+, Q_-\big\}\ &=\ 2 P\qquad\ \ \ \ \big\{\widetilde Q_+, \widetilde Q_-
\big\} \ =\ 2 \widetilde P \cr
\big\{Q_+, \widetilde Q_+\big\}\ &=\ 2 T\qquad\ \ \ \ \big\{Q_-, \widetilde Q_-
\big\} \ =\ 2 \widetilde T \ ,} }
where $P$ and $\widetilde P$ are the two light-cone components of momentum, and
$T$ and $\widetilde T$ are two central charges.   The other anti-commutators
vanish.
It is a straightforward computation using \consG\ to see that
\eqn\topch{\eqalign{
T\ \equiv\ \half \big\{Q_+, \widetilde Q_+\big\} \
& =\ - ~ \sum_i ~ t_i ~(1 - q_i) \int (dz \del_z ~+~ d\bar z \del_{\bar z})
\phi_i (z,\bar z)\cr
&=\ - \sum_i ~t_i (1 - q_i) \Big[\phi_i(\sigma = +\infty) ~-~ \phi_i (\sigma =
-\infty)\Big]\cr
&=\ \Big[W(x_a^{(\beta)})~-~ W(x_a^{(\alpha)} )\Big] \ \equiv\  \, \Delta W\ .
\cr}}
The last equality follows from the fact that \phidefn\ and \Wscale\ imply that:
\eqn\chgsimp{ - \sum_i (1 - q_i)~t_i~ \phi_i ~=~  \sum_i (1 - q_i)~t_i~
\partial_j W ~=~ W - \sum_a \omega_a x_a {\partial W \over \partial x_a} \ ,}
and the boundary conditions mean that the partial derivatives ${\partial W
\over \partial x_a}$ vanish at $\sigma = \pm \infty$.  A virtually identical
argument shows that $\widetilde T = (\Delta W)^*$.

Now define $Q = Q_+ - {\Delta W \over  \widetilde P } \widetilde Q_-$, where
$\widetilde P$ is the momentum component of the soliton in question.  Observe
that the adjoint of $Q$ is given by: $Q^\dagger = Q_- - {(\Delta W)^* \over
\widetilde P } \widetilde Q_+$.  Using \comalg\ and \topch\ in the inequality
$\big\{ Q,Q^\dagger \big\} \ge 0$ one then recovers \bogbnd, but now it has
been estalished at the quantum level.  The other thing to note is that the
bound is saturated if and only if $Q$ and $Q^\dagger$ annihilate the soliton,
that is, if and only if the soliton is {\it fundamental} ({\it i.e.} chiral).

\subsec{Examples}

I now consider what the foregoing tells us about the spectrum of two of the
quantum integrable models obtained from the $N=2$ superconformal minimal
series. In section 5, I showed that the least relevant perturbation, $\phi_n$,
gives rise to a superpotential that is the Chebyshev polynomial \Wpotb.  All
the critical points lie on the real $x$ axis, and between the $j^{\rm th}$ and
$(j+1)^{\rm th}$ critical points one has
$$
\Delta W ~=~ (-1)^{j+1} ~4 ~t^{{1 \over 2}(n+2)} \ .
$$
Consequently the fundamental solitons must run between consecutive ground
states on the real $x$-axis, and all these solitons must have the same mass.

For the most relevant, $\phi_1$, perturbation, the superpotential is given by
\Wpota.  The ground states lie at the vertices of a regular $(n+1)$-gon
centered at the origin of the complex $x$-plane, that is, at $x^{(j)} = e^{2
\pi i j \over n+1 }~t^{1 \over (n+1)}$, for $j = 0,1, \ldots n$.  Let a type
$p$ soliton be one that runs between the $j^{\rm th}$ and $(j+p)^{\rm th}$
ground states for any value of $j$. That is, a type $p$ soliton  subtends $p$
sides of the polygon.  All type $p$ solitons have the same mass, and elementary
high-school geometry shows that:
\eqn\massratio{ {m_p \over m_1} ~=~ {{sin \left( {\pi p \over n+1} \right)}
\over {sin \left({\pi \over n+1} \right)}} \ . }
These mass ratios are precisely the mass ratios that one finds in an $A_n$-Toda
theory, and so one should expect a connection with such theories.  There are
indeed such connections, and these are discussed in considerable detail in
\refs{\FLMW,\WLNW,\NemWar,\ALDNNW}.  Lest you be left with the impression that
the foregoing integrable model is the usual Toda model, I shall point out some
differences:  First, each type $p$ soliton is actually a supermultiplet of two
solitons of equal mass, and secondly, there is at least one such a
supermultiplet starting or finishing at each ground state.

Thus, the fact that we have a quantum exact \LG potential gives us all of the
soliton mass ratios.  In addition to this, the geometry of the ground states
also gives further information.  For example, if one scatters a type $p$
soliton running from the $j^{\rm th}$ ground state to the $(j+p)^{\rm th}$
ground state against a type $q$ soliton running from the $(j+p)^{\rm th}$
ground state to the $(j+p+q)^{\rm th}$ ground state, then there should be a
resonance to make a type $p+q$ soliton running from the $j^{\rm th}$ ground
state to the $(j+p+q)^{\rm th}$ ground state.  The resonant momentum can be
determined from the exact knowledge of the masses.  It turns out that in more
complicated quantum integrable models, these geometric constraints are
sufficient to determine all of the charges of all of the fundamental solitons
under all of the conserved quantities of the theory \refs{\WLNW, \WLNWB}.

\newsec{Conclusions and Apologia}

In these lectures I have not given any explanation of why certain perturbations
of $N=2$ superconformal models lead to quantum integrable theories.  I have
taken such a course, not only because of limitations in time and energy, but
also because the analysis of such issues is closely parallel to that for the
non-supersymmetric, and for the $N\! = \! 1$ supersymmetric, field theories.
The basic ideas behind this are therefore covered in Mussardo's lectures.  I
have instead chosen to stress precisely the subjects that are special to the
study of the $N=2$ supersymmetric theories, namely the exact quantum
information that can be obtained from the chiral ring and an effective \LG
potential.  Even within this restricted purview, I have omitted several very
interesting aspects of the subject.  Some of these omissions will be taken care
of by Nemeschansky and Vafa.  I have also probably made egregious errors in
referencing, for which I apologize.  What I hope to have accomplished is to
convince the reader that $N=2$ supersymmetric field theories in two dimensions
exhibit a beautiful interplay between classical and quantum structures, and
that it is this aspect of the subject that has given rise to its remarkable
vitality.

\vskip 1cm
\centerline{\bf Acknowledgements}

I am extremely grateful to my collaborators: M.~Bershadsky,  P.~Fendley,
A.~LeClair, W.~Lerche, S.~Mathur, Z.~Maassarani,  D.~Nemeschansky, D.-J.~Smit,
C.~Vafa and E.~Verlinde, without whom this work would not have been possible.
I would also like to thank the ICTP in Trieste for its hospitality and the
opportunity to organize the material presented here into a, hopefully, coherent
set of lecture notes.

\vskip 1cm
\centerline{\bf Dedication}

Brian Warr was a young English high-energy physicist who graduated from Caltech
in 1986.  He worked as a post-doctoral fellow at the University of Texas in
Austin, and also at SLAC.  In the early 1980's he contracted AIDS, and he
succumbed to it in 1992.  While Brian may not be the first physicist to die of
AIDS, he was certainly my first close friend to have been lost to this disease.
 Brian was an incandescent character who delighted in all forms of disputation,
and even during the years that he had AIDS, he radiated life.  The planet is a
poorer place for his absence.

\vfill
\eject
\listrefs
\bye
\end